\documentstyle[preprint,eqsecnum,aps]{revtex}
\input psfig.sty
\begin{document}
\tightenlines
\hyphenation{Li-brar-y}
\hyphenation{Pro-gram}
\hyphenation{U-ni-ver-si-ty}
\hyphenation{a-non-y-mous}
\title{{\tt optics}: general-purpose scintillator \\ light
response simulation code}
\author{E. Frle\v z\footnote{Corresponding author; 
Tel: +1--804--924--6786, Fax: +1--804--924--4576, 
E--mail: frlez@virginia. edu\hfill},
B. K. Wright,
D. Po\v cani\'c}
\address{Department of Physics, University of Virginia, Charlottesville,
VA 22904-4714, USA}
\maketitle
\begin{abstract}
We present the program {\tt optics} that simulates the light response of 
an arbitrarily shaped scintillation particle detector. Predicted light 
responses of pure CsI polygonal detectors, plastic scintillator staves, 
cylindrical plastic target scintillators and a Plexiglas light-distribution 
plate are illustrated. We demonstrate how different bulk and surface 
optical properties of a scintillator lead to specific volume and 
temporal light collection probability distributions. High-statistics 
{\tt optics} simulations are calibrated against the detector responses 
measured in a custom-made cosmic muon tomography apparatus. The presented 
code can also be used to track particles intersecting complex geometrical 
objects.
\par\noindent\hbox{\ }\par\noindent
{PACS Codes: 07.05.Tp; 24.10.Lx; 29.40.Mc; 87.59.F}
\par\noindent\hbox{\ }\par\noindent
{\sl Keywords:}\/ Computer modeling and simulation; Monte Carlo
simulation of scintillator response; Scintillation detectors; 
Computed tomography
\end{abstract}

\vfill\eject

\widetext

{\noindent\parindent=0pt\parskip=0.2cm
\twocolumn
\small
\centerline{\bf PROGRAM SUMMARY}
\footnotesize

{\sl Title of program}:\/ {\tt optics}

{\sl Catalogue identifier}:\/ ADxx

{\sl Program Summary URL:}\/
{\tt http:\-//www\-.cpc\-.cs\-.qub\-.ac\-.uk\-\-/cpc\-.summaries\-/ADxx} 

{\sl Program obtainable from}:\/ CPC Program Library, Queen's University
of Belfast, N. Ireland, also at 
{\tt http:\-//pi\-beta\-.phys\-.vir\-gin\-ia\-.edu\-/pub\-lic\_html\-/optics} 
or the authors.

{\sl Licensing provisions}:\/ none

{\sl Computers}:\/ Tested on MicroVAX 3100 and DECstation 5000/200. Program
should be easily portable to any UNIX workstation.

{\sl Operating system under which the program has been tested}:\/ DEC VMS 
V5.5 system in a batch mode and the DEC OSF/1 V1.3A UNIX environment in 
both a batch mode and with a graphical user interface running in a batch or 
interactive mode

{\sl Programming languages used}:\/ {\tt FORTRAN} and {\tt Tk/Tcl} toolkit

{\sl Program libraries used}:\/ CERNLIB programs in {\tt packlib} library 
and {\tt kernlib} {\tt FORTRAN} callable libraries (optional)

{\sl Memory required to execute with typical data set}:\/ up to 
1 Mb with a GUI interface

{\sl Number of bits in a word}:\/ 32 (CMOS CPU) or 64 (Alpha AXP CPU processor)

{\sl Peripherals used}:\/ standard input, standard output, hard disk,
X-terminal (optional), postscript printer for graphics 
(optional, preferably color printer)

{\sl Number of lines in the distributed programs, including test data files
and help files}:\/ 26162

{\sl Program Structure}:\/ Code consists of 65 individual files containing
the subroutines, data files and command files. A user can modify or expand
the photon transport code as well as database files specifying 
default optical properties of the detector surfaces and the bulk media.

{\sl Distribution format}:\/ uuencoded and compressed tar file or,
alternatively, compressed tar file

{\sl Keywords}:\/ Computer modeling and simulation; Monte Carlo
simulation of scintillator response; Scintillation detectors;
Computed tomography

{\sl Nature of physical problem}:\/ Simulation of the volume and temporal
light collection probability distributions given the geometrical shape plus
bulk and surface optical properties of a scintillation detector.

{\sl Method of solution}:\/ The code recognizes cylindrical, spherical, and
parabolical as well as arbitrary polygonal scintillator shapes (and 
optional wrapping reflectors) that could couple via lightguides or 
wavelength shifters to photosensitive surfaces. 
The light-generating volume can be subdivided into the elementary cells.
The photons generated within each cell are tracked through 
the scintillating volume taking into account specular, diffuse and rough surface
reflections from lateral detector surfaces and wrapping reflectors, and the bulk 
attenuation and scattering effects from detector defects~\cite{Wri92,Wri94}.

{\sl Restriction on the complexity of the problem}:\/ The statistical
uncertainties of the simulated light collection probability distribution 
are limited by the practically tolerable running time (see below).

{\sl Typical running time}:\/ The running time depends on the number
of elementary volume cells chosen and the number of scintillating 
photons generated per cell and is therefore problem-dependent. 
For example, assuming a small-step volume subdivision into a 
15$\times$15$\times$30 matrix with 6750 elements and aiming for better 
than 2\% average uncertainty in the three-dimensional light nonuniformity 
function typically requires 10$^7$ photon statistics per cell and 
running time of $\sim\,$24 hours on a 200 MHz computer.  

{\sl References}:\/ [1] B. K. Wright, Program {\tt optics}
(University of Virginia, Charlottesville, 1992).

[2] B. K. Wright, Program {\tt tkoptics} 
(University of Virginia, Charlottesville, 1994).
}
\onecolumn\normalsize

\centerline{\bf LONG WRITE-UP}

\section{Introduction}

Scintillation detectors are today widely used in nuclear and 
particle physics experiments for detection of ionizing charged 
particles as well as photons and neutrons. These detectors are 
also often used in medical instrumentation, process control
devices, waste management, personal protection and nuclear 
safeguarding~\cite{BIC}.

In the simplest design, the scintillation detector consists of
a scintillator assembly (a solid crystal, liquid or gaseous 
scintillator volume) and an optional lightguide or waveshifting 
guide viewed by a light-sensitive amplifying device such as 
a photomultiplier tube (PMT)~\cite{Leo87}.
The fraction of energy deposited in the scintillator material by
penetrating ionizing particles is absorbed and re-emitted within several 
nanoseconds in the form of visible or ultraviolet light, so-called fluorescent 
radiation. The scintillator material itself can be an organic crystal, 
plastics or liquid, inorganic crystal, scintillating gas or glass 
silicate~\cite{bir60,bir64}. Scintillation light propagates through 
the scintillator medium and bounces off its surfaces. A fraction of the light 
eventually reaches the photosensitive area of the light-sensitive device where 
it is converted into an electrical signal. The generated current pulse is typically 
discriminated in a user-designed electronics circuit and integrated for a large number 
of events to produce the pulse-height energy spectrum.

The main considerations in particle detector design are the total light
output and the detector timing response, with both factors influencing 
the energy resolution of the detector. The amount of scintillation light 
reaching the photosensitive surface should be maximized and should ideally 
be independent of the position and linearly dependent on the magnitude of 
the energy deposition. The timing response is determined by 
the decay constants of scintillating excitation and the size and geometry 
of the detector itself, as well as the characteristics of the photo-sensitive
device. The light output of a scintillator primarily 
depends on the conversion efficiency of the deposited ionizing energy to 
scintillation photons. The light collected by the photosensitive surface is 
also dependent on the surface area, the efficiency of the light transport and 
the scintillator transparency to its own scintillation light. The match
between spectral response of the photomultiplier, the scintillation light and 
the quantum efficiency of the photocathode are additional controlling factors 
determining photoelectron statistics. 

The magnitude and linearity of the pulse-heights and timing resolution of 
the scintillator signal can be optimized by tuning the light transfer to 
the photosensitive surface. The optimization is executed by the appropriate 
optical treatment of the detector surfaces and by specific choices of 
shapes and dimensions of the scintillator/lightguide assembly and 
different wrapping reflectors.

The analytical solution of the light transfer and collection process
can be obtained easily for only a few relatively simple geometries. If 
the shapes of detectors and lightguides are complex or irregular, 
or if the number of the detector modules is large, the Monte Carlo solution 
of the problem is the most practical and sometimes the only feasible 
calculational technique~\cite{jam80}. 

In order to facilitate the detector design several simple computer programs 
were developed in the past twenty-five years and were used to study 
photon transport through the elements of the scintillation 
detectors~\cite{Fal70,Mas76,Sch80,Der82,Gab87,Kno88,Hil89,Car90,Bea94}. 
These studies focused on the possibilities of improving the light collection 
and  the uniformity of light response along the main detector axes. 
Refs.~\cite{Bea94,Frl98} compared the measurements of pulse-height spectra 
of the cosmic muons and radioactive sources that were moved along 
the detector surfaces with the simulation results. The measurements of 
the scintillator spatial light output nonuniformities are reported in
Refs.~\cite{Sch87,Gri90,Dow90} without comparisons to any matching 
simulations. 

To our knowledge all published photon transport programs to date have treated 
the properties of rotationally symmetric or regular parallelepipedic detector 
shapes. The only publicly available Monte Carlo photon transport program 
that is flexible enough to treat a larger variety of detector 
shapes is the CERN Program Library {\tt GUIDE7} code~\cite{Mas76}.

We initially developed the {\tt optics} library of subroutines to study 
the light transport in simple cylindrical and trapezoidal scintillator 
detectors~\cite{Frl93}. As our need to understand the realistic photon 
propagation in the more complicated detector geometries increased, we 
expanded the code to handle a wider range of detector shapes and included 
more realistic photon--detector boundary interactions. 

In Section~\ref{sec:geo} we show how the {\tt optics} subroutines define 
the detector geometry. We explain our method of specifying the physical 
regions, region boundaries, intersections of photon trajectories with 
the region interfaces and calculation of the normal vectors at the points of 
photon--surface interactions. Section~\ref{sec:inter} touches upon the problem
of calculating the intersection coordinates between the directed lines and
the planar, cylindrical, spherical, conical and parabolical surfaces. 
Section~\ref{sec:track} presents our algorithm for the photon transport
through the detector volume. The overall program structure, including the
required and optional input parameters and a user-controlled output format 
is described in Sec.\ \ref{sec:stru}. The four progressively more 
complicated examples of the program applications are given in 
Sec.~\ref{sec:ex}. An example of using the {\tt optics} program in 
conjunction with the standard {\tt GEANT} code that simulates 
energy depositions caused by ionizing radiation in particle detectors 
is demonstrated in Sec.~\ref{sec:geant}. As an illustration of the physics 
calculation, we show how to simulate the pulse-height spectra of 
monoenergetic electrons and photons in a pure cesium iodide (CsI) 
calorimeter~\cite{Poc95}. Finally, the instructions for users outlining the 
required steps in the program installation are given in 
Sec.~\ref{sec:instal}. 

\section{Detector Geometries}\label{sec:geo}

\subsection{Region Specification}\label{sec:reg}

We define a three-dimensional region by specifying the boundaries of
the region and a point {\tt X0(3)} inside the region. A convenient way
of storing the region definitions is a {\tt FORTRAN} record structure.
While originally a record structure was a VAX {\tt FORTRAN} extension to 
the ANSI Standard, nowadays almost all {\tt FORTRAN} compilers provide
these extensions~\cite{VAX}. The record structure of
a three-dimensional region is defined in the file {\tt REGION\_STRUCTURE.TXT}:
\begin{verbatim}
C       Module REGION_STRUCTURE

C       |||||||||||||||||||||||||||||||||||||||||||||||||||||||||||||||
C               Record structure for 3-dimensional regions.
C       |||||||||||||||||||||||||||||||||||||||||||||||||||||||||||||||

        PARAMETER       (MAX_SURF = 100)

        STRUCTURE /MATERIAL_STRUCTURE/
         REAL*4         REFRACTIVE_INDEX
         REAL*4         ATTENUATION_LENGTH
         REAL*4         SCATTERING_LENGTH
        END STRUCTURE

        STRUCTURE /REGION_STRUCTURE/ 

         REAL*4 X0(3)

         INTEGER*4      NPLANE
         INTEGER*4      NCONE
         INTEGER*4      NSPHERE
         INTEGER*4      NPARABOLOID

         RECORD /PLANE_STRUCTURE/       PLANE(MAX_SURF)
         RECORD /CONE_STRUCTURE/        CONE(MAX_SURF)
         RECORD /SPHERE_STRUCTURE/      SPHERE(MAX_SURF)
         RECORD /PARABOLOID_STRUCTURE/  PARABOLOID(MAX_SURF)

         RECORD /MATERIAL_STRUCTURE/    MATERIAL

        END STRUCTURE

C       |||||||||||||||||||||||||||||||||||||||||||||||||||||||||||||||

C       End module REGION_STRUCTURE.
\end{verbatim}

The maximum number of surfaces of a specific type is declared in the user 
parameter {\tt MAX\_SURF}, while the actual number of surfaces of each type 
bounding a region is specified by the integer variables {\tt NPLANE}, 
{\tt NCONE}, {\tt NSPHERE} or {\tt NPARABOLOID}. The medium filling the region
is indexed by an integer variable {\tt MATERIAL}. For example, the assignments 
{\tt REFRACTIVE\_INDEX(1)=1.54}, {\tt REFRACTIVE\_INDEX(2)=1.00} describe 
a scintillator medium in region 1 and air in region 2, 
respectively. A database of 17 different materials commonly 
encountered in scintillation detector design is kept in the file 
{\tt MATERIAL\_PROPERTIES.TXT}.
Presently defined media properties are the light attenuation length,
scattering length, refractive index, atomic number, atomic weight,
and the material volume density. A user can add additional materials and media 
properties or change the default values of the database entries.

Once the record structure of a region is defined, any number of variables 
having this structure format can be declared. Arrays of regions can be 
defined with the statement {\tt RECORD /REGION\_STRUCTURE/ REGION(100)}
that creates an array of 100 records, each with the structure format defined 
by the module {\tt REGION\_STRUCTURE} above. The items in the structure of an
individual region are referred to as ``{\tt record.item}''. The material of
the region number 12 would therefore be referred to as 
``{\tt REGION(12).MATERIAL}'' while the $y$-coordinate of a point internal 
to the region 3 could be accessed through a variable {\tt REGION(3).X0(2)}.

Specific two-dimensional and three-dimensional shapes are defined as record 
structures in the file {\tt GEOMETRY\_STRUCTURES.TXT}. This module defines 
the structures for a point, line, circle, plane, cylinder, cone, paraboloid, 
and sphere, as well as a general polygon structure specified by a set of vertices. 
The comments in the file explain the parameters defining each shape. 
A cylindrical volume is, for example, defined by the following piece of code:

\begin{verbatim}
        STRUCTURE /CYLINDER_STRUCTURE/
         REAL*4 X0 (3)  !Point on axis of cylinder.
         REAL*4 A  (3)  !Unit vector in direction of axis.
         REAL*4 RADIUS  !Radius of cylinder.
         INTEGER*4      TYPE    !Type of surface.
         RECORD /SURFACE_PROPERTIES_STRUCTURE/ PHYSICAL
        END STRUCTURE
\end{verbatim}

Again, the user can add new geometrical shapes by defining 
additional structures conforming to the above format.

The program begins with a call to the subroutine {\tt DEFINE\_REGIONS}
where the parameters of the problem are either defined or read in from 
a user data file. An example input format for a cylindrical 
scintillator could be:
{\tt 
\begin{verbatim}
C Read in the geometry parameters  !Diameter of the front and back face
       READ(6,*) FRONT_D1, BACK_D1 !of region 1 
       READ(6,*) Z1                !z-extent of region 1

C Define material types of regions
       REGION(1).MATERIAL=2 !Scintillator
       REGION(2).MATERIAL=1 !Air

C Define interior points of regions
       REGION(1).X0(3)=Z1/2.0
       REGION(2).X0(3)=Z1+2.0

C Define boundary surfaces of regions
       REGION(1).NPLANE=2
       REGION(1).NCONE=1

C Detector surfaces
       REGION(1).PLANE(1).TYPE=1
       REGION(1).PLANE(1).N(3)=-1.0
       REGION(1).PLANE(1).X0(3)=0.0

       REGION(1).PLANE(2).TYPE=1
       REGION(1).PLANE(2).N(3)=1.0
       REGION(1).PLANE(2).X0(3)=Z1

       REGION(1).CONE(1).TYPE=1
       REGION(1).CONE(1).A(3)=1
       REGION(1).CONE(1).SLOPE=(BACK_D1-FRONT_D1)/(2.0*Z1)

C Last volume defined contains all the others
       REGION(2).NSPHERE.=1
       REGION(2).SPHERE(1).TYPE=-2
       REGION(2).SPHERE(1).RADIUS=2.0*Z1
\end{verbatim}
}

This piece of code defines a cylindrical scintillator with a {\tt z1} cm long
$z$-axis and front and back face diameters
{\tt FRONT\_D1} and {\tt FRONT\_D2}, respectively. The detector is
contained inside a spherical volume filled with air.
The different surface types are described in Sec.~\ref{sec:type}.

\subsection{Polygonal Geometries}

The irregular polygonal volume can be described by a set of vertices
stored in the {\tt POINT\_STRUCTURE} record:

\begin{verbatim}
        STRUCTURE /POLYGON_STRUCTURE/
         RECORD /POINT_STRUCTURE/ VERTEX(100)   !Set of vertices.
        END STRUCTURE
\end{verbatim}

This format gives the number of vertices of the front and back detector face 
in the first input line, and the $x,y,z$ coordinates of the front ($z=0$) 
and back detector face vertices in the second and third lines, respectively. 
For example, the truncated pentagonal detector geometry can be defined with 
a three-line entry:
{
\footnotesize\tt
\begin{verbatim}
5
2.45 3.37  0.20 3.55 0.0  0.52 7.11 0.0  0.52  8.21 3.37  0.20 5.33  5.46  0.0
0.00 6.24 22.02 2.04 0.0 22.61 8.63 0.0 22.61 10.66 6.24 22.02 5.33 10.09 21.66    
\end{verbatim}
}

The above input file describes a 22$\,$cm long truncated pentagonal pyramid with 
the front (back) face side 3.56 (6.59)$\,$cm long.
   
\subsection{Surface Specification}\label{sec:type}

The recognized types of detector surfaces are distinguished by the 
integer variable {\tt TYPE}. This parameter allows a user to specify
the physical properties of the surface. The predefined surface types 
are perfectly transparent, perfectly reflecting, and perfectly absorbing 
interfaces as well as the ``realistic'' dielectric surface. These idealized 
surfaces are distinguished in the transport code by the parameter {\tt TYPE} 
that is assigned {\tt 0, -1, -2} or positive integer values. A user should also 
set the optical properties of a surface through the structure {\tt 
SURFACE\_PROPERTIES\_STRUCTURE}:
\begin{verbatim}
        STRUCTURE /SURFACE_PROPERTIES_STRUCTURE/
         REAL*4 ROUGHNESS           !Number describing roughness.
         REAL*4 REFLECTIVITY        !Fraction of incident light reflected.
         REAL*4 DIFFUSE_FRACTION    !Diffuse_reflection/total_reflection.
        END STRUCTURE
\end{verbatim}
The {\tt REFLECTIVITY} values could range from 0.0 (a perfect absorber) to 1.0
(a perfect reflector). The {\tt DIFFUSE\_FRACTION}
coefficient determines the ratio between the specular reflection 
({\tt DIFFUSE\_FRACTION}=0.0) and the diffuse scattering processes
({\tt DIFFUSE\_FRACTION}=1.0, see Sec.~\ref{sec:diff}). The {\tt ROUGHNESS} 
parameter describes the deviation of the detector sides from an ideal flat 
surface behavior (Sec.~\ref{sec:rough}). 

It is important to note that a pair of adjacent regions that are not nested 
is always separated by the two surfaces. This arrangement allows for a
boundary between two regions to be of a different type, depending on the
region side from which a photon is approaching the interface. An example is 
a ``two-way mirror'' between two regions that could be created by making
the surface of one region perfectly reflecting and the adjacent surface
of the neighboring region completely transparent. 

The four different detector geometries are described as examples in more detail 
in Sec.~\ref{sec:ex} and shown in Fig.~\ref{shapes}. The parameters of
four geometries are specified in Table~\ref{tab1}. These shapes represent 
the real detectors whose light responses we studied both in Monte Carlo 
simulations and in calibration measurements using cosmic muons, monoenergetic 
tagged positron and photon beams, and laser pulses. The geometries are as 
varied as irregular truncated pyramids made of pure cesium iodide, plastic 
veto scintillator staves, a cylindrical plastic target, and a rectangular 
Plexiglas monitoring plate, respectively.   

\section{Intersection of Directed Lines with {\tt optics} Surfaces}
\label{sec:inter}

A straight line in three-dimensional space can be represented
by the parametric equation:
{\boldmath
\begin{equation}
X=X_{0L}+rV,
\end{equation}
}
\noindent where {\boldmath $X_{0L}$} is a point on the line,
{\boldmath $V$} is a unit vector in the direction of the line, and
the parameter {\boldmath $r$} specifies the distance from a fixed point
{\boldmath $X_{0L}$} to a point {\boldmath $X$}.

The parametric equations of the plane, spherical, and cylindrical
surfaces are given by
{\boldmath
\begin{eqnarray}
n\cdot (X-X_{0P}) &=& 0,  \text{ plane,} \\
\vert X-X_{0S}\vert &=& R, \text{ sphere,} \\
\vert X-X_{0C}-[A\cdot (X-X_{0C})] \vert &=& 0, \text{ cylinder,} 
\end{eqnarray}
}
\noindent where {\boldmath $n$} is a normal vector to the plane,
{\boldmath $X_{0S}$} and {\boldmath $R$} are the center and the radius of 
the sphere, and {\boldmath $X_{0C}$}, {\boldmath $A$} are the point
on the cylinder axis and a unit vector in the direction of the detector
$z$ axis, respectively.

Intersections of a straight line with these surfaces are found in,
for example, Ref.~\cite{Wri92}. The formulas for the intersection 
coordinates are implemented in the {\tt optics} subroutine 
{\tt SUBROUTINE FIND\_LINE\_REGION\_INTERSECTION(LINE,I\_REGION,
X1,X2,INTERSECT,REGION,N\_REGIONS)}.
The relevant regions should first be defined with calls to 
{\tt SUBROUTINE DEFINE\_REGIONS\_CRYSTAL(REGION,N\_REGIONS)}.

The {\tt LINE} variable is specified by the three-dimensional array 
{\tt LINE.X0} and by the three cosines of directions  {\tt LINE.V} that 
could be calculated from the polar and azimuthal angles ($\theta$,$\phi$):
\begin{eqnarray}
{\tt LINE.V(1)}&=&\sin\theta\cos\phi, \\
{\tt LINE.V(2)}&=&\sin\theta\sin\phi, \\
{\tt LINE.V(3)}&=&\cos\theta.
\end{eqnarray}
The variable {\tt I\_REGION} is an index of a region that takes values from 
1 to {\tt N\_REGIONS}, while
the {\tt REGION} variable is the {\tt FORTRAN} record structure described 
in Sec.~\ref{sec:reg}.

{\tt INTERSECT} is a logical variable that is {\tt .TRUE.} if at least  
one intersection point is found and {\tt .FALSE.} if there are no 
intersections between the line and user defined surfaces. The arguments 
{\tt X1} and {\tt X2} are three element arrays that contain
the calculated coordinates of the intersection points.

In Fig.~\ref{hexa1} we show the intersections of reconstructed cosmic
muon trajectories with the surfaces of a cesium iodide
scintillation crystal. The crystal's geometrical shape is
a truncated irregular hexagon labeled HEX--A (Fig.~\ref{shapes}.i). 
Fig.~\ref{plastic1} illustrates the cosmic muon trajectories intersecting 
three vertically stacked rectangular scintillator bars 
(Fig.~\ref{shapes}.ii). In both cases the intersection coordinates have been
calculated using the {\tt FIND\_LINE\_REGION\_INTERSECTION} subroutine.
\bigskip

\section{Photon Tracking}\label{sec:track}
\subsection{Dielectric Reflection}
The reflection and refraction of light from a perfect dielectric interface 
can described by the laws of geometrical optics if the scintillation photons
are unpolarized. If the angles between the directions of the incident, 
reflected and transmitted photon with respect to the surface normal are 
denoted with $\theta_i$, $\theta_r$ and $\theta_t$, respectively, the law 
of reflection states that
\begin{equation}
\theta_r=\pi-\theta_i,
\end{equation}
while the law of refraction (Snell's law) on the other hand requires that
\begin{equation}
n_1\sin\theta_i=n_2\sin\theta_t,
\end{equation}
\noindent where $n_1$ and $n_2$ are the indices of refraction of the media 
in regions 1 and 2, respectively.

The user can change the program so that the full polarization of the
propagating photon is taken into account. The reflection and transmission
phenomena would then be described by the Fresnel formulas
(see, for example,~\cite{Fal70,Jac75}).

\subsection{Diffuse Reflection}\label{sec:diff}

A perfectly diffuse reflector is defined by the uniform scattering
probability $P(\theta)$ into a unit solid angle $d\Omega$, where $\theta$
denotes the angle between the direction of a reflected photon and the normal
vector to the surface. The angular distribution of the scattered radiation 
is then isotropic. If the reflecting surface is a plane, the resulting
distribution is often called the Lambert's (or cosine) law~\cite{Bor93}:
\begin{equation}
I(\theta)=I_0\cos\theta.
\end{equation}

\subsection{Rough Surfaces}\label{sec:rough}
A rough surface can be modeled by introducing a set of small plane
surfaces tangent to the original surface at suitable points. This description
goes under the name of the facet model and is described in more detail
in Ref.~\cite{Bea94} where the model predictions are also compared with
the experimental measurements.

In the simplest version of the facet model a rough surface is characterized 
by a single roughness parameter $R$. This 
parameter is zero for smooth scintillator surfaces with a perfect finish. 
For real detectors with uneven boundaries $R$\/ value is a positive number 
between 0.0 and 1.0. Polished rough surfaces with height variations of
the order of a micron have the roughness coefficient on the order of 0.1.

The subroutine {\tt PERTURB\_UNIT\_VECTOR
(V\_IN,ROUGHNESS,V\_OUT)} calculates a random perturbation to the direction of
an incident photon unit vector {\tt V\_IN} following the reflection from 
a rough surface. 
The scalar product of the perturbed vector {\tt V\_OUT} and the unperturbed 
photon direction $\cos\theta$ is equal to $1/\sqrt{(1+\delta^2)}$, where 
$\delta$ has a Gaussian distribution, centered at zero, with the width given 
by the {\tt ROUGHNESS} parameter $R$.

\subsection{Absorption and Rescattering}
The media properties are described by two parameters, 
{\tt SCATTERING\_LENGTH} and {\tt ATTENUATION\_LENGTH}. 
The code simplifies photon transport by assuming that the scattering length 
and the absorption length do not depend on the wavelength of the 
propagating light. The dispersion effects are neglected. If a user wants
to consider dispersion effects, the changes should be
made in the file {\tt optics.f}. 
\bigskip

\section{Program Structure}\label{sec:stru}
\subsection{Program Input}\label{sec:input}

The program starts with the subroutine {\tt DEFINE\_REGIONS} 
that reads a data file specifying the parameters of the 
detector. The call to that subroutine defines the record structures
in the array {\tt REGION}. A starting position of a scintillation photon 
is accepted next. The coordinate values can be read in from the data file 
{\tt PHOTONS.DAT} or, alternatively, a user can select one of 
the predefined distributions:
\begin{enumerate} 
\item a single point source, fixed $x,y,z$;
\item a through-going particle, range of $x$, fixed $z$, $y=0$;
\item a range of light sources on the detector axis, range of $z$, $x=y=0$;
\item a uniform light distribution throughout detector volume,
range of $x,y,z$;
\item a uniformly distributed  light sources in the center plane, range 
of $x,z$, $y=0$;
\item an aimed ``pencil'' of scintillating radiation, fixed $x,y,z$, 
and fixed directional cosines $v_x,v_y,v_z$;
\item a uniform distribution in the first half (first half-sector) of 
the detector, range of $x,y,z$.
\end{enumerate}

All input parameters defining the detector and the scintillating
light distribution, as well as the desired output format can also be 
entered from the {\tt tkoptics} window if one is using the GUI
version of the program (see Fig.~\ref{fig:tkoptics}).
\bigskip
\subsection{Physics Subroutines}

The initial region number is found next by calling the routine
{\tt FIND\_REGION\_NUMBER}. The program then steps through
the following levels:
\medskip

1. Find the next intersection of a photon trajectory with the predefined 
   surface.
\smallskip

2. Test to see if a photon is absorbed or rescattered in transit.
\smallskip

 2.1 If absorption or rescattering occurred, look at the current 
     region number.
\smallskip
     
    2.1.1 If a photon is in the wavelength shifter, find the
    point at which it is absorbed, generate a new random direction, 
    and continue.
\smallskip

    2.1.2 If a photon is not in the wavelength shifter, count it
    as absorbed and get a new particle.
\smallskip
   
 2.2 If a photon is not absorbed, continue.
\smallskip

3. Propagate a photon to the next intersection.
\smallskip

4. Look into the next region to find its region number.
\smallskip

 4.1 If the next region number is the same as the current region
     number, continue the photon transport and return to (1).
\smallskip

 4.2 If we are dealing with a different region, continue to (5).
\smallskip

5. Look at the type of the surface a photon encounters.
\smallskip

 5.1 If the surface is transparent, continue to the next intersection.
\smallskip

 5.2 If the surface is reflecting, stay in the current region while
     changing the photon's direction.
\smallskip

 5.3 If the surface is absorbing, stop the photon propagation, count it as
     absorbed, and generate a new photon ab initio and return to step (1).
\smallskip

 5.4 When an interface is a real surface, use the Snell's law to determine
     if the photon undergoes reflection or refraction, change 
     the direction appropriately, and step into the next region 
     if a photon is refracted.
\smallskip

 5.5 For the watch points, call the {\tt WATCHPOINT} subroutine, then do (5.1).
\smallskip

 5.6 For the detectors, call the {\tt DETECTOR} subroutine, count a photon 
     as detected, then do (5.3).
\smallskip

6. Return to (1).
\medskip

This algorithm is represented by a box diagram in Fig.~\ref{fig:algorithm}.

The following is the list of subroutines used in the photon propagation:
{\tt
\begin{verbatim}
SUBROUTINE INITIALIZE ( REGION, N_REGIONS )
SUBROUTINE WRITE_STATS ( REGION, N_REGIONS )
SUBROUTINE GENERATE_PHOTON ( LINE, REGION, N_REGIONS )
SUBROUTINE LOG_PHOTON_GENERATION ( LINE )
SUBROUTINE LOG_INTERSECTION ( LINE, REGION, N_REGIONS, N_CURRENT, 
                              R, X, OUTWARD_NORMAL, CLASS, ISURF, TYPE )
SUBROUTINE LOG_REFLECTION ( LINE, REGION, N_REGIONS, N_CURRENT,
                            R, X, OUTWARD_NORMAL, CLASS, ISURF, TYPE )
SUBROUTINE LOG_REFRACTION ( LINE, REGION, N_REGIONS, N_CURRENT, 
                            R, X, OUTWARD_NORMAL, CLASS, ISURF, TYPE )
SUBROUTINE LOG_ABSORPTION ( LINE, REGION, N_REGIONS, N_CURRENT, 
                            R, X, OUTWARD_NORMAL, CLASS, ISURF, TYPE )
SUBROUTINE LOG_ATTENUATION ( REGION, N_CURRENT, LINE, X )
SUBROUTINE DEFINE_REGION_HALF_SECTOR ( FILENAME, 
SUBROUTINE DEFINE_REGIONS ( REGION, N_REGIONS )
SUBROUTINE DEFINE_REGIONS_PYRAMID ( REGION, N_REGIONS )
SUBROUTINE DEFINE_REGIONS_CONE ( REGION, N_REGIONS )
SUBROUTINE DETECTOR ( TYPE, LINE, RANGE, TIME )
SUBROUTINE WATCHPOINT ( TYPE, LINE, RANGE, TIME )
SUBROUTINE TERMINATE ( REGION, N_REGIONS )
\end{verbatim}
}

The functions these subroutines perform are described in 
Sec.~\ref{sec:track}.

\subsection{Program Input}\label{inp}
The listing of the program input for the graphical user interface (GUI) 
version of the code is printed by default at the program termination:
{\tt
\begin{verbatim}
 Enter geometry type: 1
 	0 = Truncated cone
 	1 = Pi-Beta CsI crystal
 Enter pyramid type: 0
 	1	= Pentagon
 	2	= Hexagon-A
 	3	= Hexagon-B
 	4	= Hexagon-C
 	5	= Hexagon-D
 	6	= Half Hexagon-D (#1)
 	7	= Half Hexagon-D (#2)
 	0	= User defined (will prompt for file name)
 Enter file name: /fs24/users/emil/tomo/tkoptics/hh1-d-mc.dat
 Enter refractive index of pyramid material: 2.10
 Enter attenuation length of pyramid material (cm): 200.0
 Enter scattering length of pyramid material (cm): 200.0
 Enter surface type for sides of pyramid: -1
 	 1	= normal dielectric
 	-1	= imperfect specular reflector
 	-2	= perfect absorber
 	-3	= imperfect diffuse reflector
 Enter reflectivity of this surface (0-1): 0.975
 Enter roughness of this surface (0-1): 0.0
 Enter diffuse fraction for this surface (0-1): 0.0 
 Enter surface type for front face of pyramid: -1
 	 1	= normal dielectric
 	-1	= imperfect specular reflector
 	-2	= perfect absorber
 	-3	= imperfect diffuse reflector
 Enter reflectivity of this surface (0-1): 0.975
 Enter roughness of this surface (0-1): 0.0
 Enter diffuse fraction for this surface (0-1): 0.0 
 Enter refractive index of air gap material: 1.000273
 Enter attenuation length of air gap material (cm): 1.0E+3
 Enter scattering length of air gap material (cm): 1.0E+3
 Enter surface type for pyramid wrapper: -3
 	 1	= normal dielectric
 	-1	= imperfect specular reflector
 	-2	= perfect absorber
 	-3	= imperfect diffuse reflector
 Enter reflectivity of this surface (0-1): 0.9
 Enter roughness of this surface (0-1): 0.0
 Enter diffuse fraction for this surface (0-1): 1.0 
 Enter surface type for front face of wrapper: -3
 	 1	= normal dielectric
 	-1	= imperfect specular reflector
 	-2	= perfect absorber
 	-3	= imperfect diffuse reflector
 Enter reflectivity of this surface (0-1): 0.2
 Enter roughness of this surface (0-1): 0.0
 Enter diffuse fraction for this surface (0-1): 1.0 
 Enter refractive index of optical joint material: 1.58
 Enter attenuation length of optical joint material (cm): 200.0
 Enter scattering length of optical joint material (cm): 200.0
 Enter refractive index of PMT window material: 1.458
 Enter attenuation length of PMT window material (cm): 100.0
 Enter scattering length of PMT window material (cm): 100.0
 ********* Xfront =   6.790000       2.865000       22.19750    
 ********* AXIS =  0.0000000E+00  0.1182974      0.9929782    
 ********* VECTOR =  8.4076812E-03 -5.3805402E-03 -6.0056243E-04
 ********* VECTOR =  0.0000000E+00  1.1829736E-02  9.9297822E-02
 ********* Xfront =   6.790000       2.876830       22.29680    
 Enter output data type: 3
   0 = No output 
   1 = Endpoint data
   2 = Tracks
   3 = Statistics
 Enter output file name: /fs24/users/emil/tomo/tkoptics/tkoptics.rz
 Enter numbers of bins: (NXBINS NYBINS NZBINS) 14 12 24
 Enter X limits of container volume: (MIN MAX) 0 14
 Enter Y limits of container volume: (MIN MAX) 0 12
 Enter Z limits of container volume: (MIN MAX) 0 24
 Enter starting distribution: 3
      0 = Single point (Fixed X,Y,Z)
      1 = Through-going particle (Range of X; fixed Z; Y=0)
      2 = Range of points along axis (Range of Z; X=Y=0)
      3 = Uniform throughout volume (Range of X,Y,Z)
      4 = Uniform in center plane (Range of X,Z; Y=0)
      5 = Aimed (Fixed X,Y,Z; Fixed VX,VY,VZ)
      6 = Uniform in first half-sector (Range of X,Y,Z)
 Enter number of photons: 5000000
 Number of photons processed =      5000000
  
 Number of photons attenuated =      1467024
 Number of photons absorbed =      2197182
 Number of photons overbounced =      169244
  
 Number of photons detected =     1166550
 	(      489568 started backward )
 	(      676982 started forward )

\end{verbatim}
}

This output is an example of the {\tt optics} simulation of 
a half-hexagonal CsI detector, Fig.~\ref{shapes}.i.

\subsection{Program Output}

The user can choose three different output formats.
In the {\tt ENDPOINTS} format for each detected scintillation photon 
a record is written to an ASCII output file documenting:
\begin{enumerate}
\item $x_0,y_0,z_0$, the starting coordinates of a generated photon;
\item $v_{x_0},v_{y_0},v_{z_0}$, the components of the initial
directional vector of the photon;
\item $x_1,y_1,z_1$, a point at which the photon hit the
photosensitive surface;
\item $v_{x_1},v_{y_1},v_{z_1}$, the components of the photon's 
direction vector when it hit the photosensitive surface;
\item $t$, the time elapsed from the photon creation to the moment
of its detection;
\item $n$, the number of times a photon was reflected from the
detector surfaces.
\end{enumerate}

The program terminates by printing the photon transport statistics summary. 
The summary includes the number of detected photons, as well as fractions of
attenuated, absorbed and ``overbounced'' photons (see the tail of 
Sec.~\ref{inp} listing). The limit to the number of photon bounces is set
by a user. The default limit is 200 reflections. Events exceeding that 
limit are counted as overbounced and the propagation
resumes with a new photon at the starting position.

The {\tt CERNLIB PAW} macro program titled {\tt tkpawread.kumac}
can be used to read in the ASCII file produced by the {\tt optics} code
and make an {\tt HBOOK4 RZ} file containing a summary Ntuple~\cite{paw}. 
The usage syntax is {\tt exec tkpawread {\sl infile outfile}}. 
The structure of the generated Ntuple is given in an example below:

{
\begin{verbatim}
          ********************************************************
          * NTUPLE ID=   10  ENTRIES= 233209   Endpoint Data     *
          ********************************************************
          *  Var numb  *   Name    *    Lower     *    Upper     *
          ********************************************************
          *      1     * xstart    * 0.180037E+00 * 0.138114E+02 *
          *      2     * ystart    * 0.146866E-03 * 0.585641E+01 *
          *      3     * zstart    * 0.561714E-02 * 0.225345E+02 *
          *      4     * vxstart   * -.999965E+00 * 0.999996E+00 *
          *      5     * vystart   * -.999988E+00 * 0.999991E+00 *
          *      6     * vzstart   * -.999990E+00 * 0.999998E+00 *
          *      7     * xstop     * 0.314337E+01 * 0.115353E+02 *
          *      8     * ystop     * 0.238158E+00 * 0.541021E+01 *
          *      9     * zstop     * 0.220957E+02 * 0.227119E+02 *
          *     10     * vxstop    * -.999665E+00 * 0.998784E+00 *
          *     11     * vystop    * -.990139E+00 * 0.999993E+00 *
          *     12     * vzstop    * -.994538E+00 * 0.999994E+00 *
          *     13     * time      * 0.976813E-02 * 0.302494E+02 *
          *     14     * bounces   * 0.000000E+00 * 0.109000E+03 *
          ********************************************************
\end{verbatim}
}
\noindent The variable names correspond to the ASCII file variables.
The number of Ntuple entries in this case is equal to the number of 
generated photons.

If a user wants to calculate the three-dimensional light collection 
probability function the {\tt STATISTICS} output format is more appropriate.
The variables written to an ASCII file and into the resulting Ntuple are:
\begin{enumerate}
\item {\tt x,y,z}, the coordinates of the centers of the elementary
cells into which the detector volume is subdivided;
\item {\tt counts}, the number of detected photons originating from
a given elementary cell;
\item {\tt edge}, the flag indicating if an elementary cell is fully
contained inside the detector volume ({\tt edge$=$1}), or is partly outside 
the detector ({\tt edge$=$0});
\item {\tt thrown}, the number of the generated photons having the origins 
within a given elementary cell;
\item {\tt time}, the average time-of-flight of the photon (time elapsed 
between the generation and detection moment).
\end{enumerate}

\begin{verbatim}
          ********************************************************
          * NTUPLE ID=   10  ENTRIES=   4032   Statistics        *
          ********************************************************
          *  Var numb  *   Name    *    Lower     *    Upper     *
          ********************************************************
          *      1     * x         * 0.500000E+00 * 0.135000E+02 *
          *      2     * y         * 0.500000E+00 * 0.115000E+02 *
          *      3     * z         * 0.500000E+00 * 0.235000E+02 *
          *      4     * counts    * 0.000000E+00 * 0.957900E+04 *
          *      5     * edge      * 0.000000E+00 * 0.100000E+01 *
          *      6     * thrown    * 0.000000E+00 * 0.407280E+05 *
          *      7     * time      * 0.000000E+00 * 0.463954E+01 *
          ********************************************************
\end{verbatim}
The number of the Ntuple entries represents the number of elementary cells,
and in the case above corresponds to the 14$\times$12$\times$24$\,$cm$^3$ 
subdivision. The simulated detector was irregular trapezoidal pyramid
shown in Fig.~\ref{shapes}.i. The resulting histograms of the number of 
photon reflections as well as the distribution of the photon time-of-flight 
times are shown in Fig.~\ref{trans}. Converting the later histogram into 
a cumulative representation we get the intrinsic contribution of the photon 
arrival times to the timing lineshape of the detector system, Fig.~\ref{times}. 

Finally, the information on the individual photon trajectories could be 
captured with the {\tt TRACK} output format:
\begin{verbatim}
 ******************************************************************
 * Ntuple ID = 10     Entries = 53424     Track Data
 *************************************************************************
 * Var numb * Type * Packing *    Range     *  Block   *  Name           *
 *************************************************************************
 *      1   * I*4  *         * [0,200]      * TRACK    * ntrack          *
 *      2   * I*4  *         *              * TRACK    * itrack(ntrack)  *
 *      3   * R*4  *         *              * TRACK    * xtrack(ntrack)  *
 *      4   * R*4  *         *              * TRACK    * ytrack(ntrack)  *
 *      5   * R*4  *         *              * TRACK    * ztrack(ntrack)  *
 *      6   * R*4  *         *              * TRACK    * vxtrack(ntrack) *
 *      7   * R*4  *         *              * TRACK    * vytrack(ntrack) *
 *      8   * R*4  *         *              * TRACK    * vztrack(ntrack) *
 *************************************************************************
 *  Block   *  Entries  * Unpacked * Packed *   Packing Factor           *
 *************************************************************************
 * TRACK    *  53424    * 5604     * Var.   *    Variable                *
 * Total    *    ---    * 5604     * Var.   *    Variable                *
 *************************************************************************
 * Blocks = 1            Variables = 8       Max. Columns = 1401         *
 *************************************************************************
\end{verbatim}
The track number, starting and endpoint coordinates and the initial track
direction of a photon are recorded in this file. Figure~\ref{pmt} 
illustrates the uniform distribution of the detected photons on the surface 
of a photocathode. The simulated detector was again the HEX-D1/2 shape
shown in Fig.~\ref{shapes}.i. 
\bigskip
\section{Program Examples}\label{sec:ex}

The coordinate system is always defined with the $z$ axis pointing along 
the detector axis and the origin is placed at the front detector face. 
If the detector shape is polygonal, one vertex of the front detector face 
should always lie on the positive $x$ axis.  All detector surfaces in 
examples below are normal dielectric interfaces. Some surfaces are
wrapped in the specularly reflecting layer of aluminized Mylar or
multiple layers of Teflon sheet.
User-defined geometrical shapes are included in an encompassing sphere
filled with air. The rectangular subvolume of the sphere is divided into 
a number of elementary cells with side lengths $L$, usually $\sim$1$\,$cm.

\subsection{Irregular Pentagonal, Hexagonal and Trapezoidal Scintillators}

The coordinate system is as follows: the $z$ direction points along the axis 
of the crystal, with $z=0$ at the smaller (front) face, and $z=L$ at the
larger (back) face. One vertex of the front face and one of the back face 
lie on the positive $x$ axis. The coordinates of the vertices for each 
crystal type are stored in the text data files. The nine different 
scintillator shapes are examined: four irregular hexagonal truncated 
pyramids (we label them HEX--A, HEX--B, HEX--C, and HEX--D), one regular 
pentagonal (PENT) and two irregular half-hexagonal truncated pyramids 
(HEX--D1 and HEX--D2), and two trapezohedrons (VET--1, VET--2)~\cite{Poc95}. 
The volumes of studied CsI crystals varied from 797$\,$cm$^3$ (HEX--D1/2) 
to 1718$\,$cm$^3$ (HEX--C), see Fig.~\ref{shapes}.i. The size of the
elementary cell was 1$\,$cm$^3$.

The geometry of the simulation is simple. The CsI scintillator crystal is 
wrapped in a layer of aluminized Mylar and attached via a 0.2$\,$mm thick 
optical joint to a 1$\,$mm thick photomultiplier window. Behind this window is
a photocathode. The diameter of the photosensitive surface is 46$\,$mm (HEX--D and
VET) or 67$\,$mm inches (PENT and HEX) . The entire detector is enclosed in 
a large absorbing sphere filled with air. The program output is written to 
a binary {\tt RZ} file. The results of the calculation are the 
three-dimensional light collection probability functions illustrated on 
Figs.~\ref{fig:hexa_9} and \ref{fig:hh1d_9}. The percentage probability
that photons generated within two-dimensional $x$--$z$ 
bins will generate a photoelectron on the PMT cathode is shown on
top panels. The bottom panels are tomography measurements of the corresponding
light responses for two representative HEX--A and HEX--D1 detectors.

\medskip
\subsection{Rectangular Plastic Veto}
The cylindrical plastic hodoscope of the PIBETA detector~\cite{Poc95} used 
in charged particle tracking consists of 20 rectangular staves, 
Fig~\ref{shapes}.ii. The individual sections are optically isolated from each other 
by one layer of aluminized Mylar wrapping and are viewed by one inch phototubes 
coupled via lightguides on both ends. The simulated light response of the individual 
PMTs as a function of the axial position along the plastic stave is an example 
of the calculation of physical interest.

\medskip
\subsection{Cylindrical Plastic Target}

A plastic scintillator cylindrical target is used in the center of
the PIBETA detector~\cite{Poc95} to stop a pion beam (Fig.~\ref{shapes}.iii).
The positrons from pion and muon decays emanating from the target center 
are detected subsequently in the plastic veto hodoscope and the CsI 
calorimeter. A typical {\tt optics} calculation of the photon transport 
predicts the light response of the target as a function of the polar
angle of the exiting positron.    

\medskip
\subsection{Light-Distribution Plexiglas Plate}

The lead glass calorimeter monitor for the RadPhi experiment E94-026 
at JLAB~\cite{radphi} is a laser-based calorimeter monitoring system.
The nitrogen laser excites a cylindrical plastic scintillator and a
generated light is distributed via six fiber optics cables to the
1.2$\,$cm thick Plexiglas sheet~\cite{Ple} facing the lead glass wall in a 
light-tight enclosure (Fig~\ref{shapes}.iv). The sheet is oversized at 
152$\,$cm$\times$152$\,$cm to minimize the optical nonuniformities.
The two-dimensional response of the smaller prototype system predicted with 
the {\tt optics} code, as viewed through lead glass modules is displayed on 
Fig.~\ref{monitor}. The maximum predicted variation in the light output of 
the plate as a function of the position of an individual lead glass 
detector is $\sim$30$\,$\%. 
\bigskip

\section{Energy Lineshape Monte Carlo}\label{sec:geant}

\subsection{Interpolation of 3D Light Nonuniformity Function}

The program {\tt optics} can be used to calculate the light 
collection probability function $f(iL,jL,kL)$, where $i,j,k$ are integers 
and $L$ the step size, on a user-defined three-dimensional grid.

Starting from this three-dimensional probability function the code can calculate
for an arbitrary point source of light $(x,y,z)$ inside the detector volume
the light fraction reaching the photosensitive device. This is accomplished
by means of a series of simple one-dimensional interpolations.
Our method of choice is the {\sl bilinear 
interpolation} on an elementary grid square. The accuracy will be improved 
if one uses not only the light output values but also the gradients and
the cross derivatives of the light output probability changes.
These higher order methods that result in a smoother behavior of
our function go under the names of the {\sl bi-cubic interpolations} or
{\sl bi-cubic splines}. Numerical recipes for these methods are
given as {\tt FORTRAN} subroutines and {\tt C} functions in Ref.~\cite{Pre86,Vet86}.
\medskip

\subsection{{\tt GEANT} Simulation of Electromagnetic Showers}
The three-dimensional distribution of energy deposited in the
detector volumes by ionizing charged particles was calculated
by the standard code {\tt GEANT}~\cite{brun87}. We considered monoenergetic 
positrons and photons of 70$\,$MeV total energy incident on the PIBETA
calorimeter made of pure CsI pyramids. In the user-written
subroutine the energy deposited in every step was multiplied
by the normalized value of the light collection nonuniformity 
function to yield the detected energy deposition. Sums are kept of 
both the total deposited energy and total detected light. 
Examples of predicted energy deposition lineshapes in the
calorimeter are shown in Fig.~\ref{lineshapes}.
Four different simulations are plotted. The top panel shows a
pulse-height spectrum for $10^5$ simulated photons incident on the CsI 
calorimeter from its center. The responses of both idealized uniform detector 
and the detector characterized with the {\tt optics}-calculated 
nonuniformity functions are displayed. The bottom panel shows the difference 
in the response for the 70$\,$MeV positron showers.

\bigskip
\section{Installation and Testing}\label{sec:instal}
\bigskip
The computer code described in this paper is distributed in the form
of uuencoded compressed {\tt tar} archives named {\tt optics.uu} (229 kb) 
and {\tt tkoptics.uu} (2.4 Mb). These two files can be obtained
via transfers from the URL 
{\tt http:\-//\-pibeta\-.phys\-.virginia\-.edu\-\-/public\_html\-/optics}. Same files can
also be obtained directly from the authors. The UNIX utilities for
the {\tt tar} archive processing can be copied from the official GNU site at
{\tt http://www.gnu.ai.mit.edu}.

Installation of the {\tt optics} software on a computer running 
the UNIX operating system is straightforward. Execution of the {\tt csh} 
script {\tt uudecode} taking {\tt optics} or {\tt tkoptics} as a parameter 
creates a compressed tar archive file with an extension {\tt tar.gz}. 
The archive can be unpacked with the command {\tt gunzip}. The unpacked 
{\tt optics} library files should be placed under the {\tt 
/common\-/local\-/optics} subdirectory. The extracted {\tt tkoptics} files should 
be copied to the {\tt optics} subdirectory on the user's disk.  If the user 
prefers the GUI version of the program, the installation of 
{\tt CERNLIB} libraries, as well as of the {\tt Tk} and {\tt Tcl} 
libraries~\cite{Ous94} is required. The publicly available distributions 
can be accessed, for example, at {\tt ftp:\-//ftp\-.scriptics\-.com\-/pub\-/tcl/}. 
The parameter file {\tt .tkoptics} defining the input parameters echoed 
in the default optics window should be placed in the login directory. 

The {\tt Tk} and {\tt Tcl} software should be installed under 
{\tt /common\-/local\-/tk} and {\tt /common\-/local\-/tcl} areas, respectively. 
The most recent CERN Program Library release containing subroutines 
used in the {\tt tkoptics} program is available at the site {\tt 
ftp:\-//asisftp\-.cern\-.ch}.

The working area can be now named by the project's title and should contain 
the required input files, like the {\tt optics.dat} and {\tt tkoptics.dat}. 
To verify that the program executes properly the user should remake the 
executable code by compiling and linking the source code with the
{\tt make optics} command . The {\tt Makefile} file is included in 
the distribution. The program can then be started with the {\tt tkoptics} 
command invoking a {\tt GUI} interface window or it can be run in batch 
mode with the {\tt optics} command.

\section{Acknowledgements}
The authors wish to thank Micheal Sadler of the Abilene Christian
University for the loan of the drift chamber tomography apparatus. 
Stefan Ritt and Penelope Slocum of the University of Virginia and
David Lawrence of the Arizona State University have helped with 
the cosmic muon tomography measurements. Their help is gratefully 
acknowledged.

This work is supported and made possible by grants from the US National
Science Foundation and the Paul Scherrer Institute.
\bigskip

\vfill
\eject

\vfill
\eject
\begin{table}
\caption{Sample ranges of input optical parameters for {\tt optics}
simulations of various detectors. We simulated the response of the
hexagonal and half-hexagonal pure CsI scintillators, a plastic scintillator
stave, a cylindrical active plastic target, and a Plexiglas light
distribution plate.}
\label{tab1}
\bigskip
\begin{tabular}{lllll}
Parameter &CsI Detector &Plastic Hodoscope &Plastic Target& 1.3$\,$cm 
Plexiglas \\ 
\ &HEX--A,D1/2 &4.1$\times$0.3$\times$59.8$\,$cm$^3$ 
&\o{4.0}$\times$5.0$\,$cm$^2$& 100$\times$185$\,$cm$^2$ \\ 
\hline
Matrix Size& 12$\times$14$\times$24& 4$\times$1$\times$60& 
   10$\times$10$\times$10& 1$\times$3$\times$10 \\
Detector Surface Type& specular& specular& specular& specular\\
Detector Surface Roughness& 0.0--1.0& 0.0--1.0& 0.0--1.0&
0.0--0.5 \\
Wrapper Surface Type& Teflon& Al Mylar& Al Mylar& Al Mylar \\
Photocathode Size (cm)& \o{4.5}/\o{7.0}& \o{1.8}& \o{1.0}&
\o{3.0}\\
Photocath. QE (\%)/Timing (ns)& 20/0.5& 28/0.4& 20/0.5& 13/2 \\
Bulk Attenuation Length (cm)& 50--300& 100-400& 100-400&
100--400\\
Bulk Scattering Length (cm)& 50--300& 100-400& 100--400&
100--400 \\
Index of Refraction (blue light)& 2.08& 1.58& 1.58& 1.58 \\
Specular Reflectivity& 0.7--1.0& 0.7--1.0& 0.7--1.0& 0.7--1.0 \\
Wrapper Reflectivity& 0.5--1.0& 0.5--1.0& 0.5--1.0& 0.5--1.0 \\
\end{tabular}
\end{table}
\vfill
\eject

\begin{figure}
\caption{Some of the particle detectors and light distribution systems studied
with the {\tt optics} code: (i) HEX--A pure CsI detector, (ii) plastic
scintillator veto stave, (iii) cylindrical active plastic target, and
(iv) Plexiglas light-distribution monitoring plate. 
}
\label{shapes}
\end{figure}

\begin{figure}
\caption{The HEX--A scintillator volume intersected by the cosmic muons 
tracked using a triple drift chamber system. The physical edges of the detector volume 
are cut away by the requirement that the cosmic muon pathlengths be longer than 
1$\,$mm. The intersection points 
are calculated with {\tt FIND\_LINE\_REGION\_INTERSECTION} subroutine.}
\label{hexa1}
\end{figure}

\begin{figure}
\caption{Three plastic veto scintillator volumes intersected by tracked
cosmic muons. The top and bottom stave surfaces, separated by 0.318$\,$cm,
are outlined by the cosmic muon intersections. The vertical edges of the 
staves are missing due to the requirement that the cosmic muon pathlengths be
longer than 1$\,$mm.}
\label{plastic1}
\end{figure}

\begin{figure}
\caption{Main {\tt tkoptics} X-Window-based input menu prompts a user 
for a series of detector parameters. Predefined geometries are
conical, pentagonal, hexagonal and general polygonal detector shapes.
Detector boundary surfaces and wrapping materials could be specified
inside this window.} 
\label{fig:tkoptics}
\end{figure}

\begin{figure}
\caption{Block diagram of the {\tt optics} code logic. The starting photon 
positions and directions are read from an ASCII file or simulated according 
to a user prescription. The numbers of absorbed, ``over-bounced'' and 
detected photons are counted.}
\label{fig:algorithm}
\end{figure}

\begin{figure}
\caption{{\tt optics} simulation of the photon transport 
in the PIBETA HEX-D1 trapezoidal detector. The number of 
detected photon reflections from the detector surfaces 
and well as the arrival times of the detected photons are
shown.
}
\label{trans}
\end{figure}

\begin{figure}
\caption{{\tt optics} simulation of the scintillator
pulse shape for the PIBETA HEX-D1 trapezoidal detector.
Only the photon time-of-flight contribution to the pulse-height
shape is shown. 
}
\label{times}
\end{figure}

\begin{figure}
\caption{{\tt optics} simulation of a two inch photocathode surface
illuminated by scintillation light. The scintillation photons
were generated uniformly throughout the volume of the 
HEX-D1 trapezoidal detector, see Fig.~\protect\ref{shapes} and 
Table~\protect\ref{tab1}.
}
\label{pmt}
\end{figure}

\begin{figure}
\caption{The simulated (i) and measured (ii) light collection probabilities 
are shown as functions of axial ($z$) and transverse ($x$) coordinates for an
ideal hexagonal detector (HEX--A) with a two-layer Teflon wrapping. The size 
of the histogrammed two-dimensional bins is 1$\times$1$\,$cm$^2$. The average 
light collection probability with a 78$\,$mm$\,\phi$ inch phototube is 23$\,$\%.}
\label{fig:hexa_9}
\end{figure}

\begin{figure}
\caption{ (i) Calculated Monte Carlo light collection probability as a two dimensional
function of axial and transverse coordinates for an ideal half-hexagonal
detector (PIBETA HEX--D1) with a Teflon wrapping. The average light 
collection probability with a 46$\,$mm$\,\phi$ photocathode is 11$\,$\%. (ii) Second 
panel shows the measured response of the detector.}
\label{fig:hh1d_9}
\end{figure}



\begin{figure}
\caption{The two-dimensional light response of the RADPHI lead glass
calorimeter monitoring plate. The nitrogen laser light is fed into 
the plate via six 1$\,$mm thick optical fibers at the top and bottom sides.} 
\label{monitor}
\end{figure}

\begin{figure}
\caption{(i) Monte Carlo pulse-height spectra of 70$\,$MeV photons in 
the pure CsI modular calorimeter. The dashed line histogram represents 
the case of an ideal detector with uniform light collection probabilities. 
The full lineshape shows the energy spectrum calculated using {\tt GEANT} 
code and the light collection probabilities calculated with {\tt optics}
program (see Figs.~\protect\ref{fig:hexa_9} and \protect\ref{fig:hh1d_9}). 
(ii) The identical calculation done for 70$\,$MeV monoenergetic positrons.}
\label{lineshapes}
\end{figure}

\clearpage

\vfill
\eject


\newpage

{

\bigskip
\begin{figure}
\noindent (i)
\hbox to \textwidth
{\vbox{\hsize=2.8in
\centerline{\psfig{figure=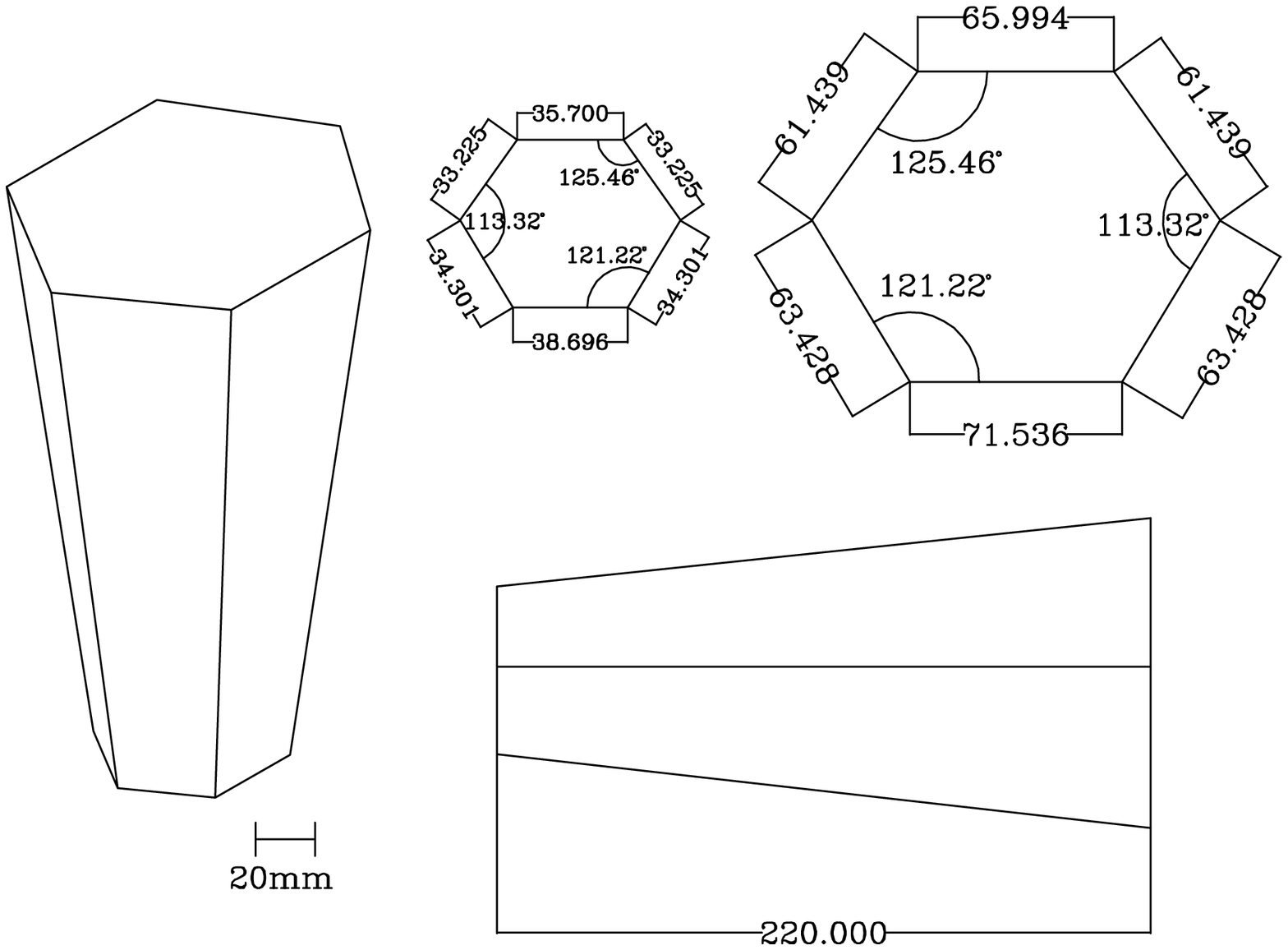,width=2.8in}}
\vglue -2.5cm
       }\hfill
\label{fig:hexa}
\vbox{\hsize=2.8in
\centerline{\psfig{figure=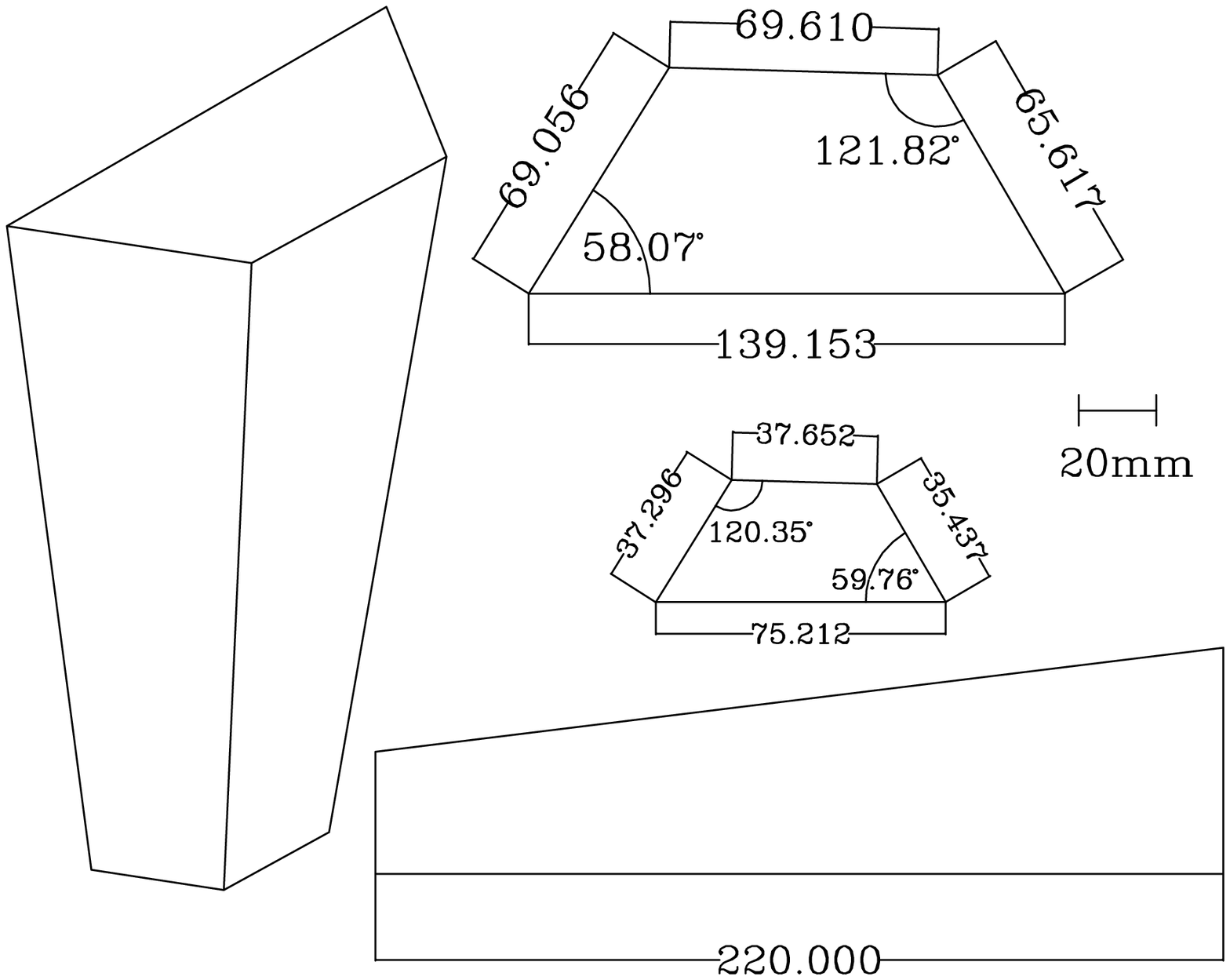,width=2.8in}}
\vglue -2.5cm
}
}
\label{fig:hexd}
\end{figure}
\bigskip
\vglue 1.8cm
\hrule

\bigskip
\begin{figure}
\noindent (ii)
\vglue -1cm
\hbox to \textwidth
{\vbox{\hsize=1.3in
\centerline{\psfig{figure=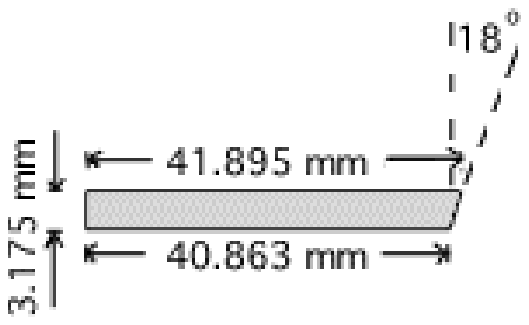,width=1.3in}}
\vglue -2.5cm
       }\hfill
\label{fig:stave_end}
\vbox{\hsize=3.9in
\centerline{\psfig{figure=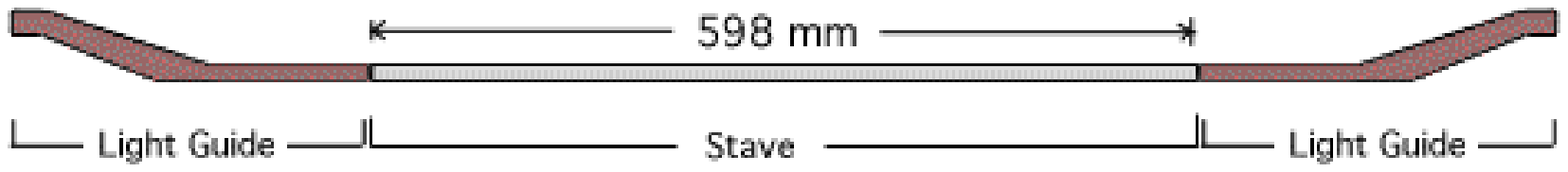,width=3.9in}}
\vglue -2.0cm
}
}
\label{fig:stave_side}
\end{figure}
\bigskip
\vglue 1.8cm
\hrule
\bigskip\bigskip
\noindent (iii)
\vglue -0.5cm
\centerline{\psfig{figure=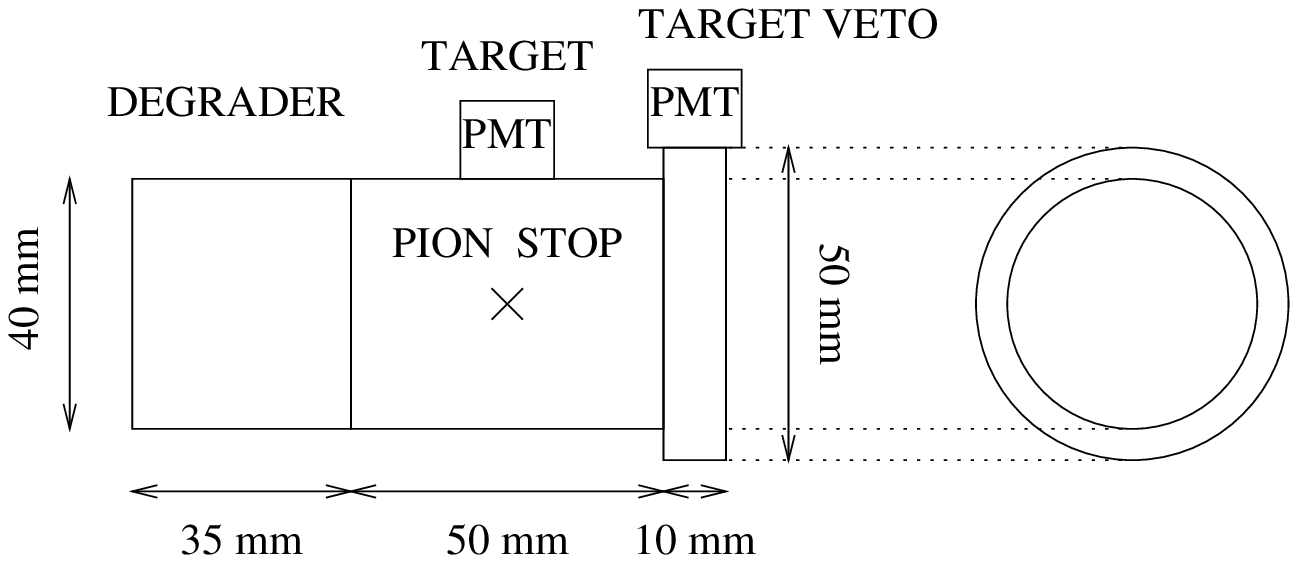,height=3.5cm}}
\vglue 1cm
\hrule
\bigskip\bigskip
\noindent (iv)
\vglue -3.5cm
\centerline{\psfig{figure=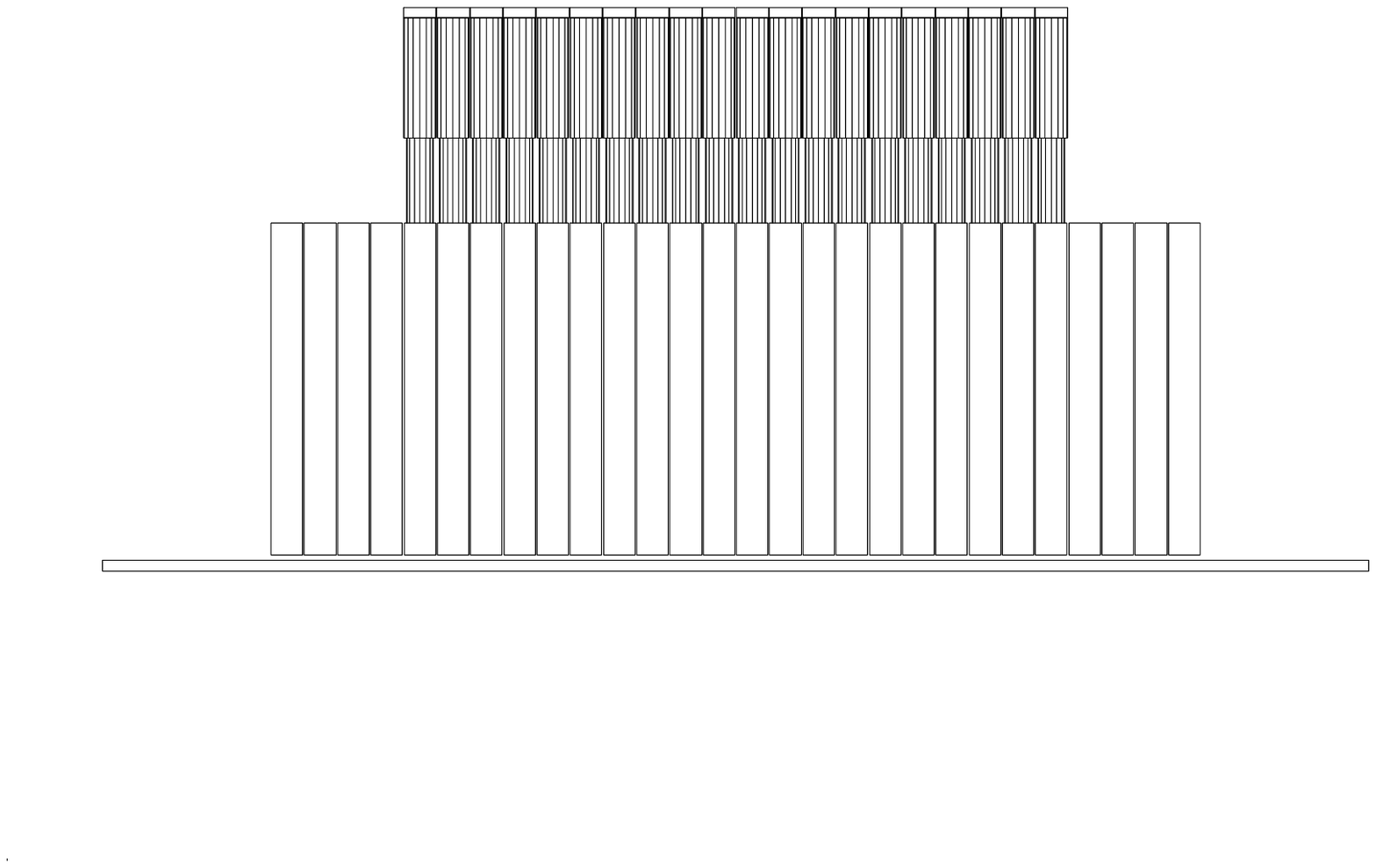,height=9.8cm}}
\vglue -2.5cm\hrule
\bigskip\bigskip\bigskip
\centerline{FIGURE 1}
\vfill\eject

\centerline{\psfig{figure=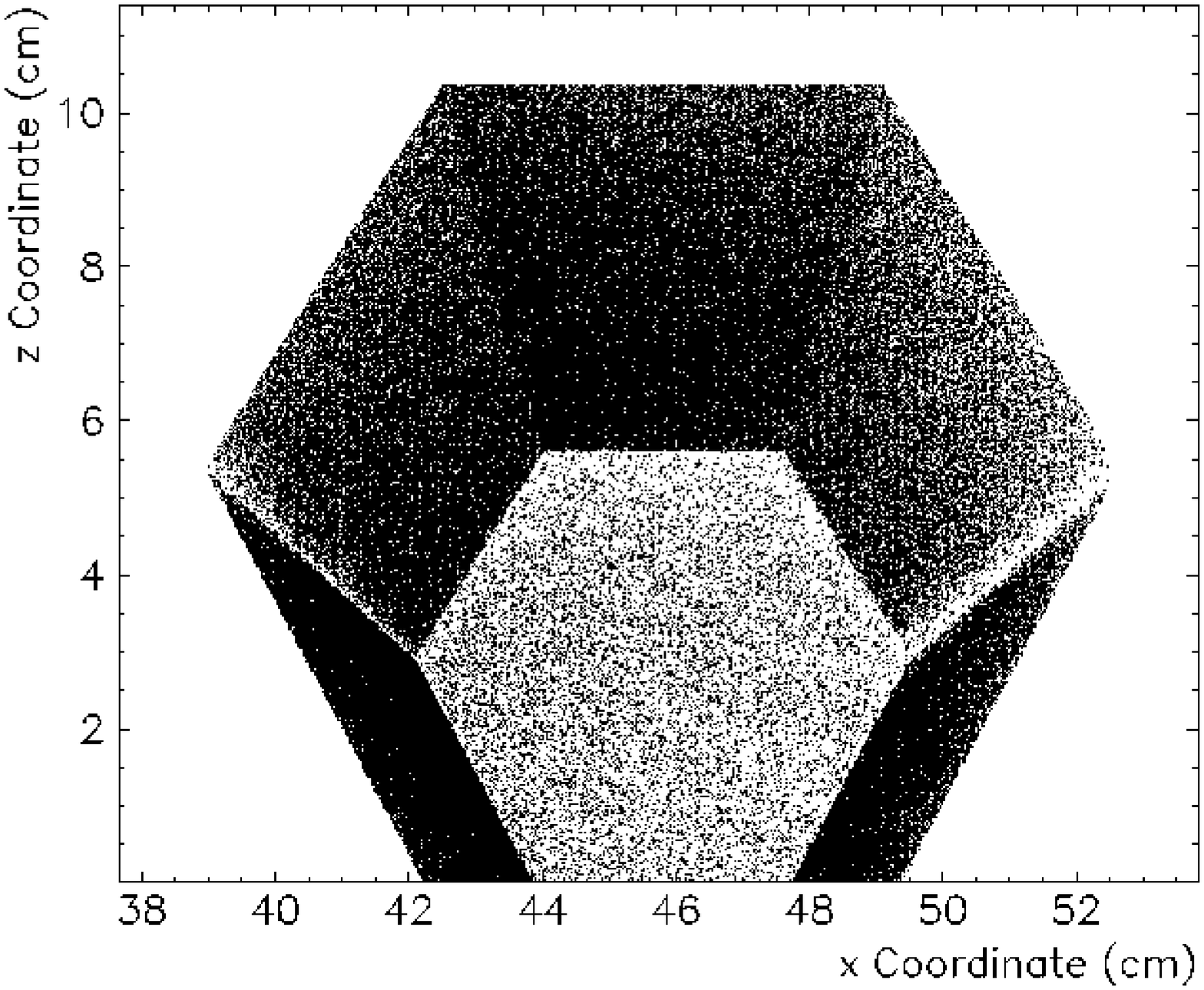,height=14cm}}
\bigskip
\centerline{FIGURE 2} 
\vfill\eject

\centerline{\psfig{figure=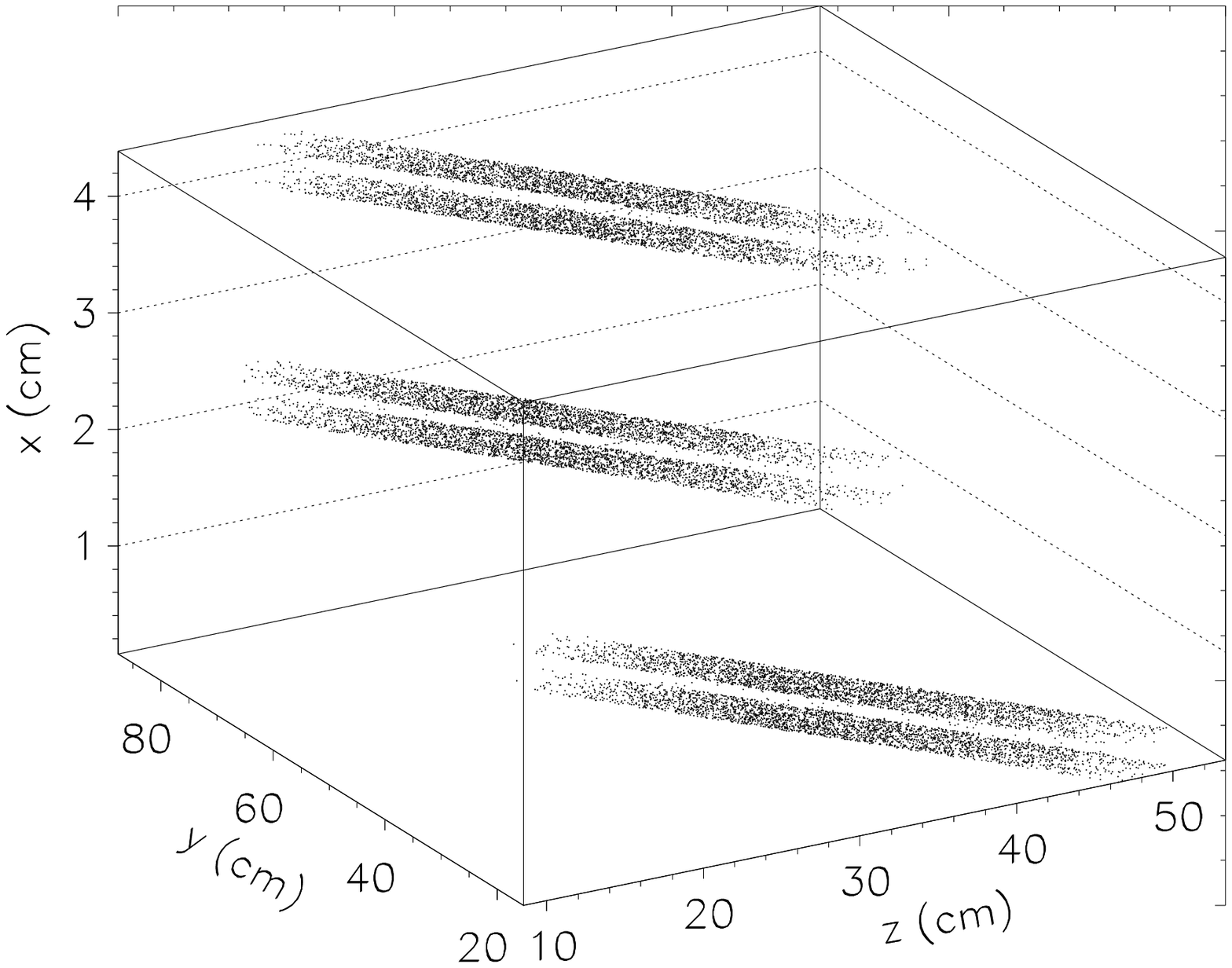,height=14cm}}
\bigskip
\centerline{FIGURE 3} 
\vfill\eject

\centerline{\psfig{figure=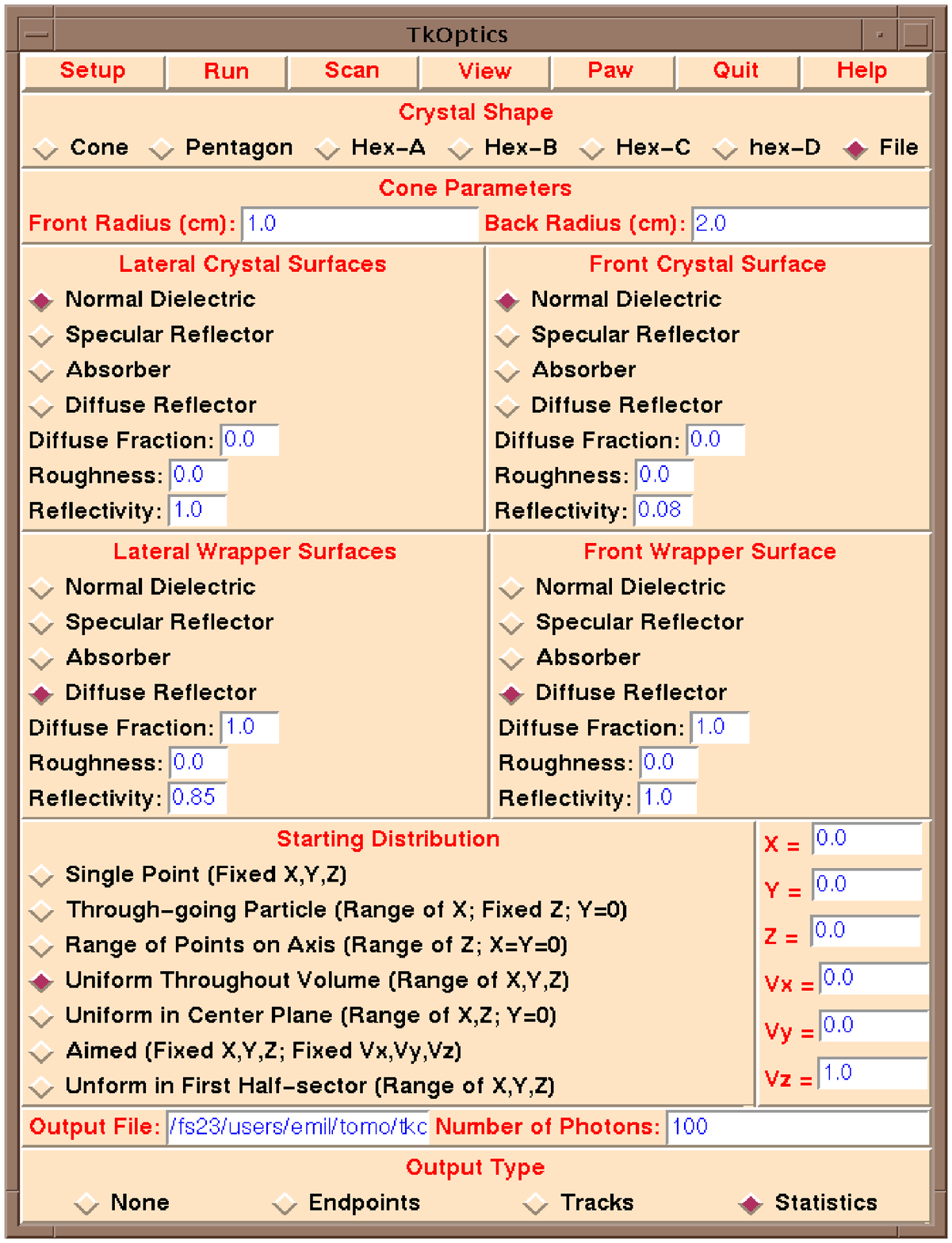,height=20cm}}
\vglue -0.5cm
\centerline{FIGURE 4} 
\vfill\eject 

\centerline{\psfig{figure=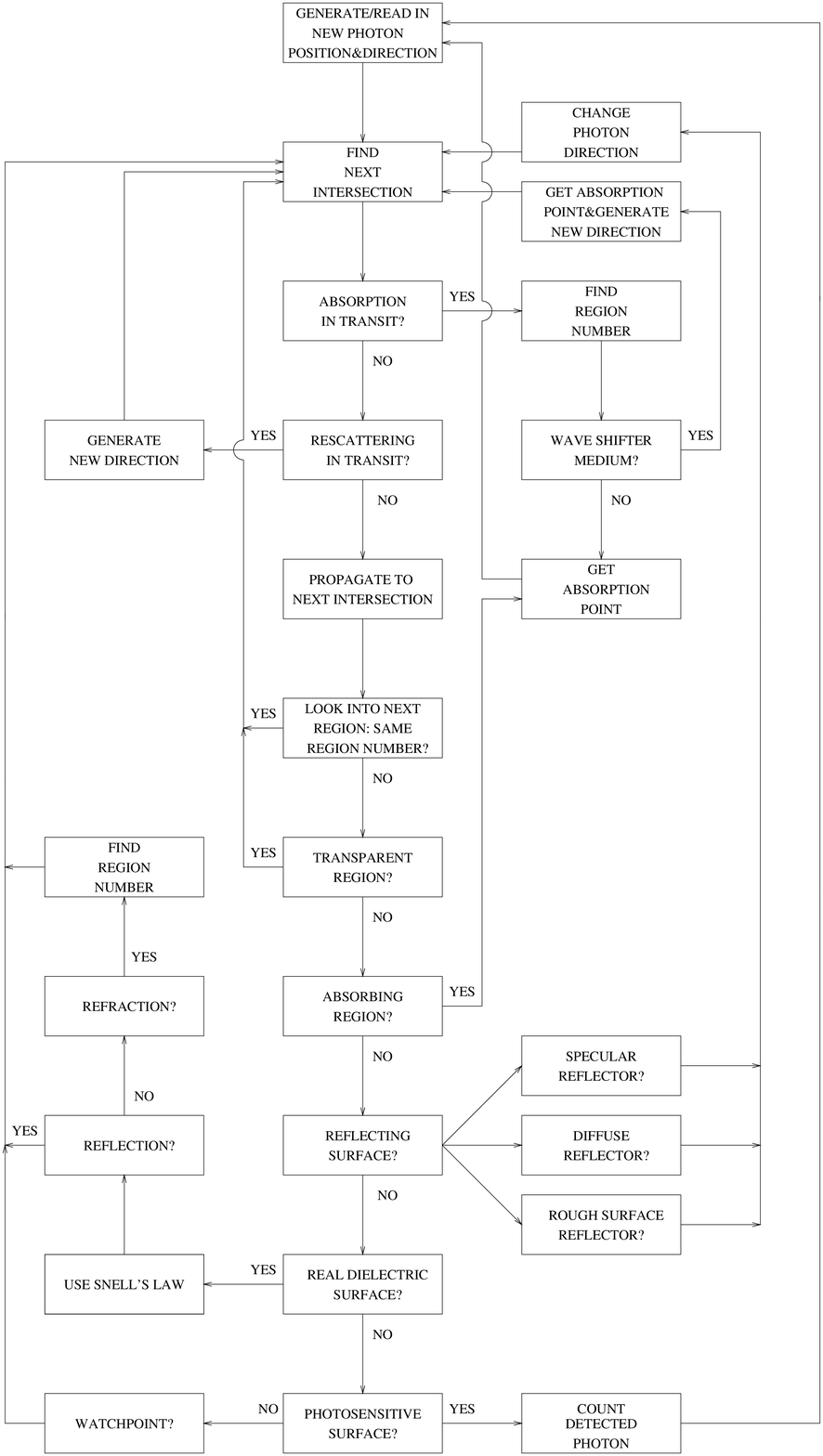,height=20cm}}
\bigskip\bigskip\bigskip
\centerline{FIGURE 5} 
\vfill\eject  

\centerline{\psfig{figure=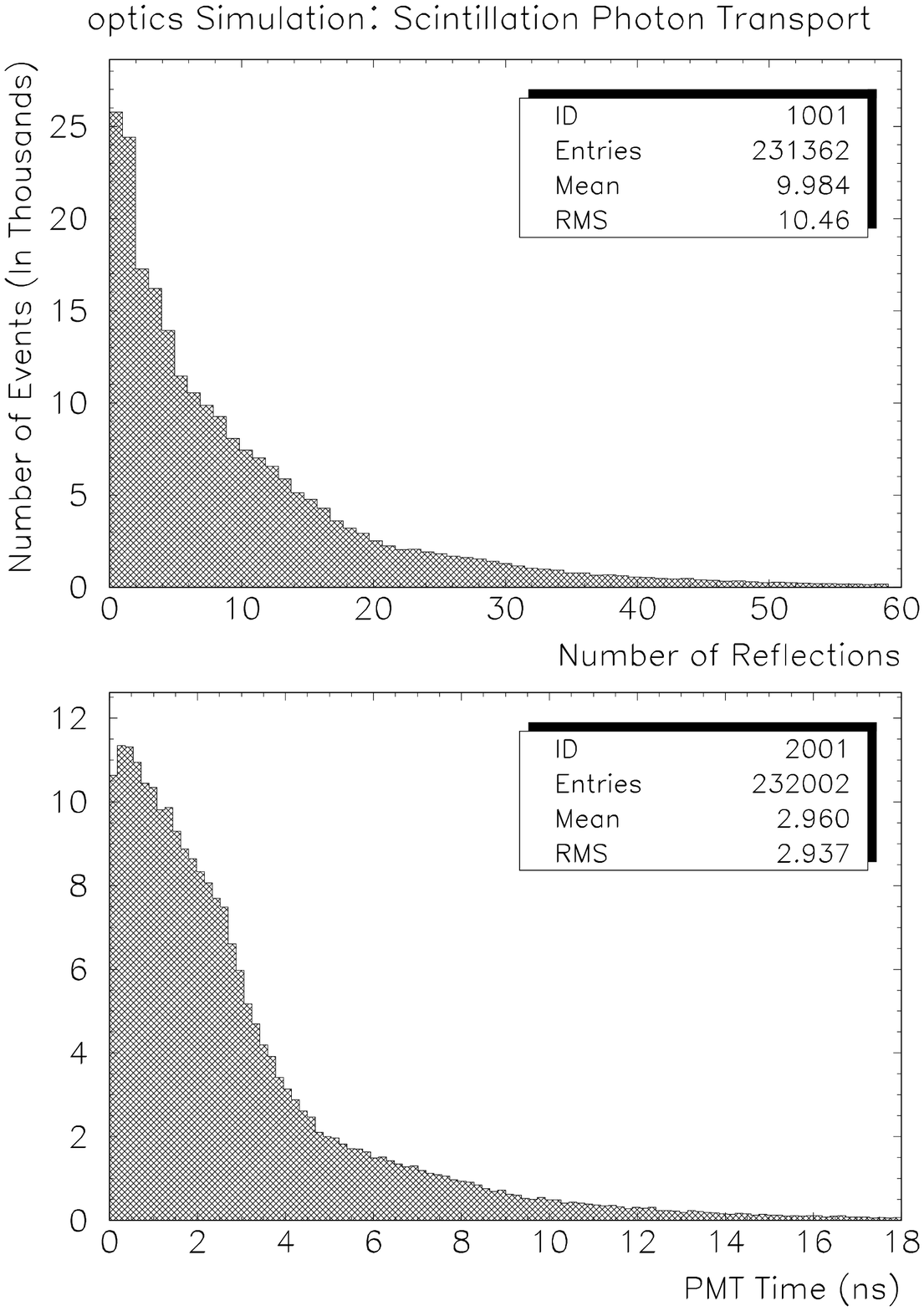,height=20cm}}
\bigskip\bigskip
\centerline{FIGURE 6}
\vfill\eject

\centerline{\psfig{figure=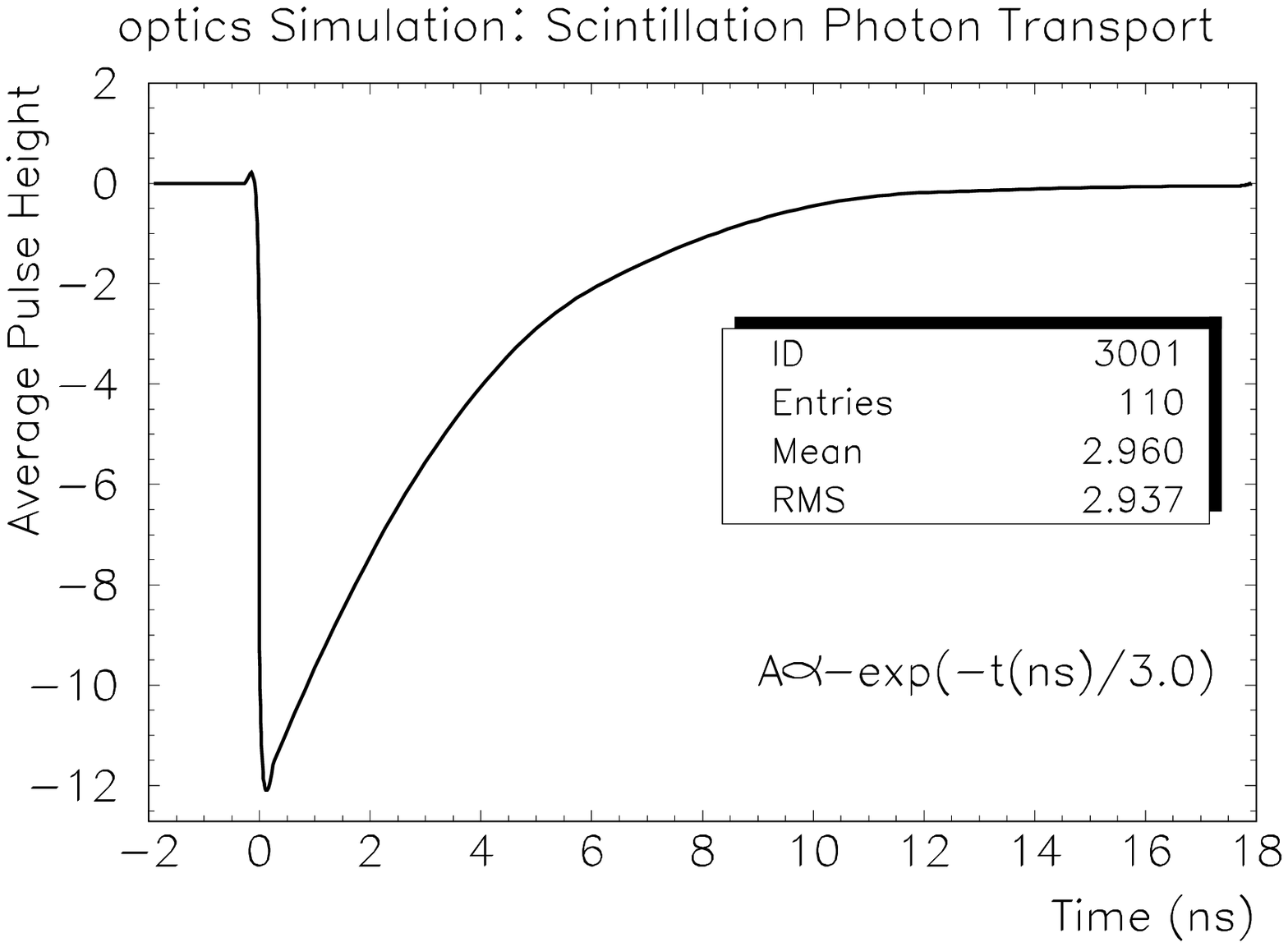,height=20cm}}
\vglue -8.0cm
\centerline{FIGURE 7}
\vfill\eject

\centerline{\psfig{figure=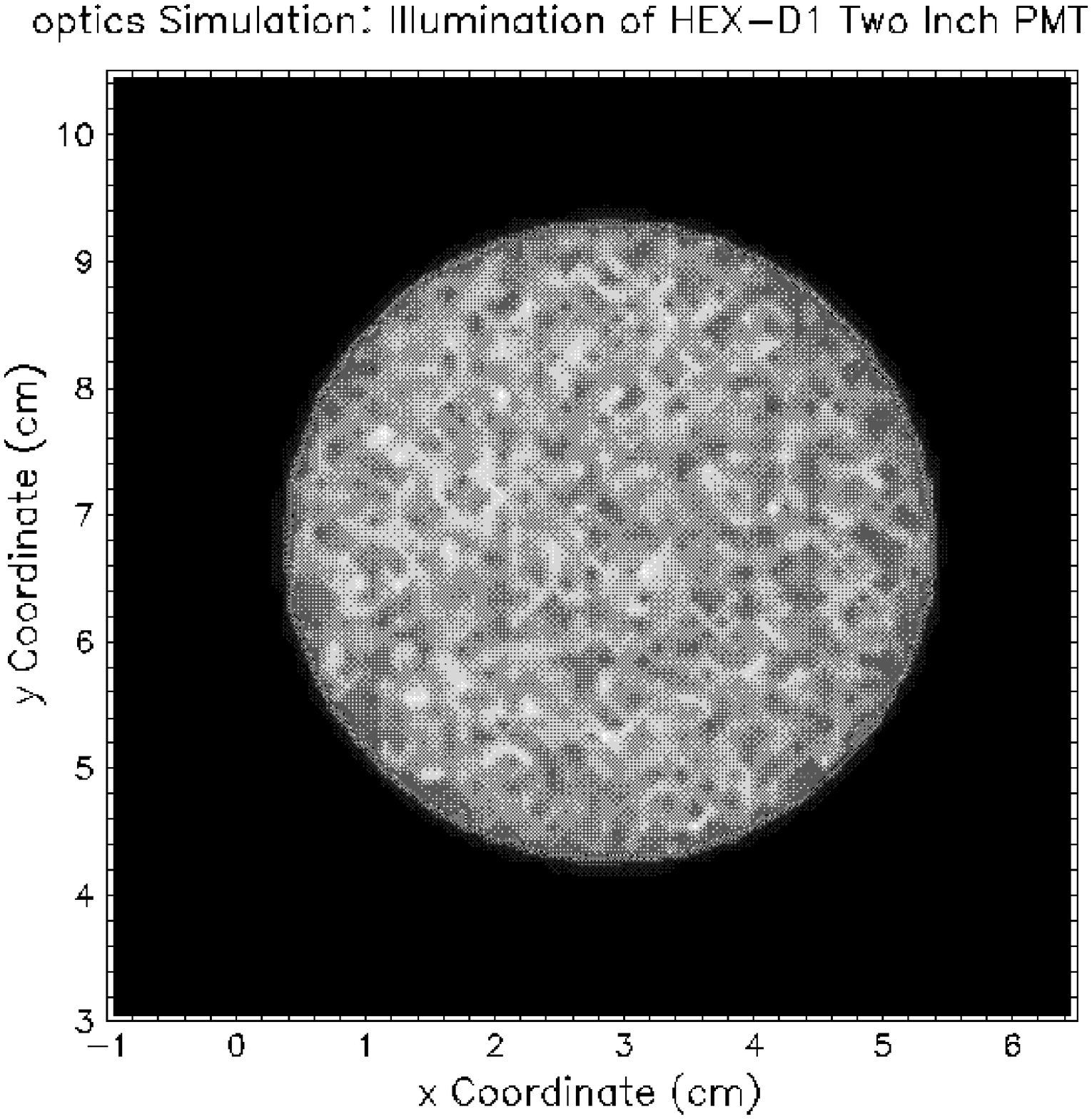,height=14cm}}
\bigskip\bigskip
\centerline{FIGURE 8}
\vfill\eject

\noindent (i)\vglue -2.5cm
\centerline{\psfig{figure=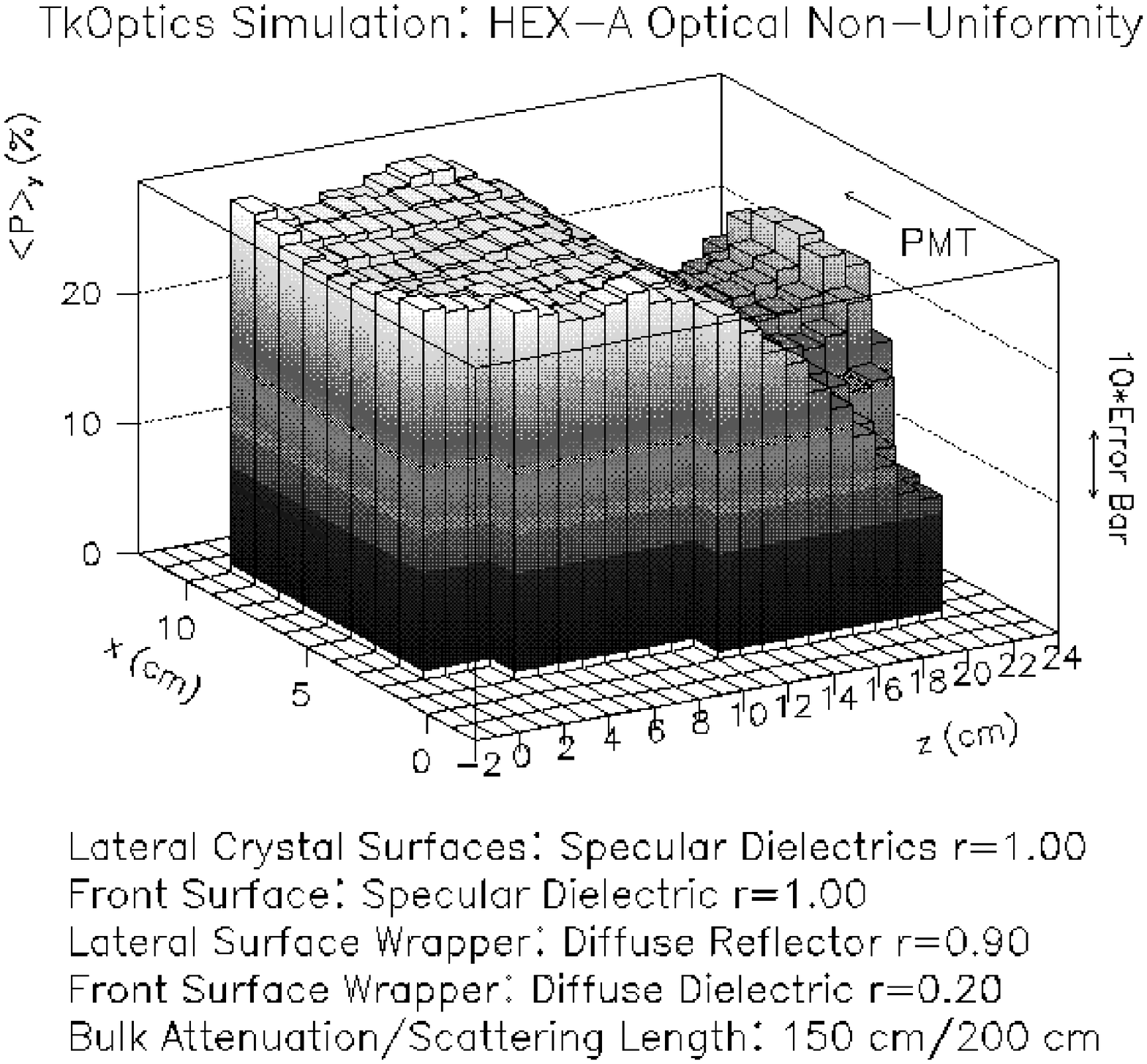,height=13.8cm}}
\vglue -2.0cm
\noindent (ii)
\centerline{\psfig{figure=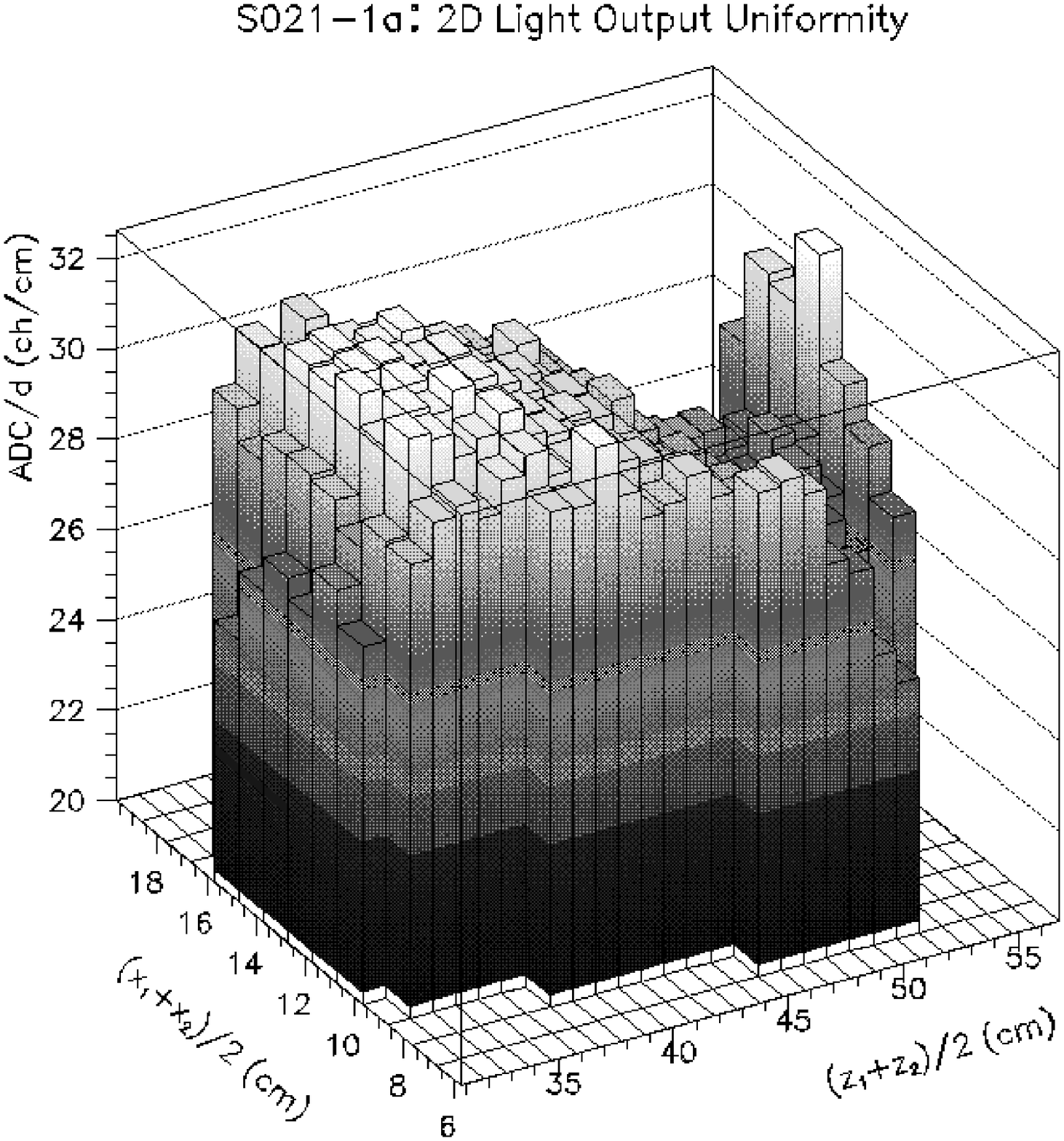,height=11.7cm}}
\centerline{FIGURE 9} 
\vfill\eject

\noindent (i)\vglue -2.5cm
\centerline{\psfig{figure=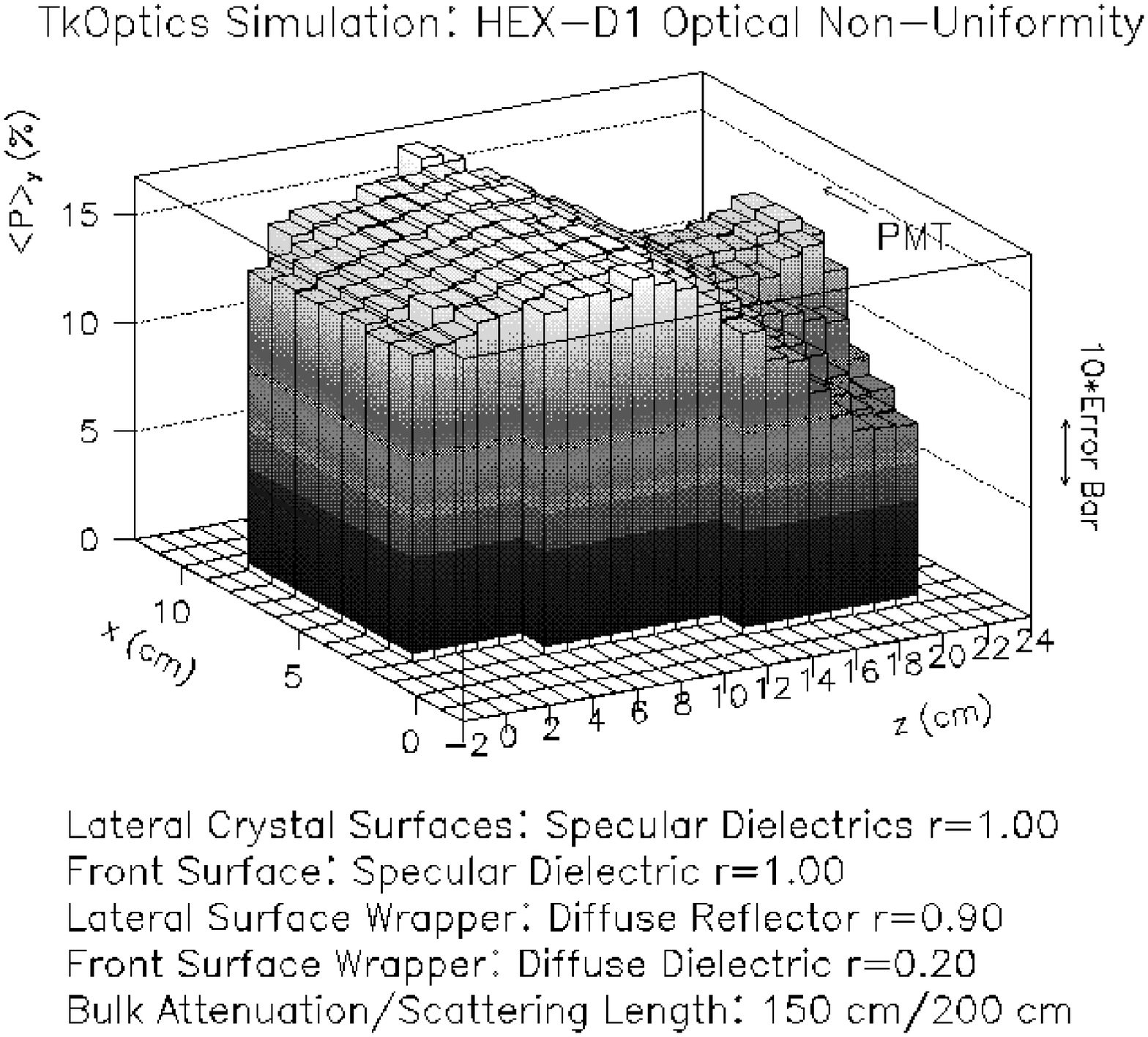,height=13.8cm}}
\vglue -2.0cm
\noindent (ii)
\centerline{\psfig{figure=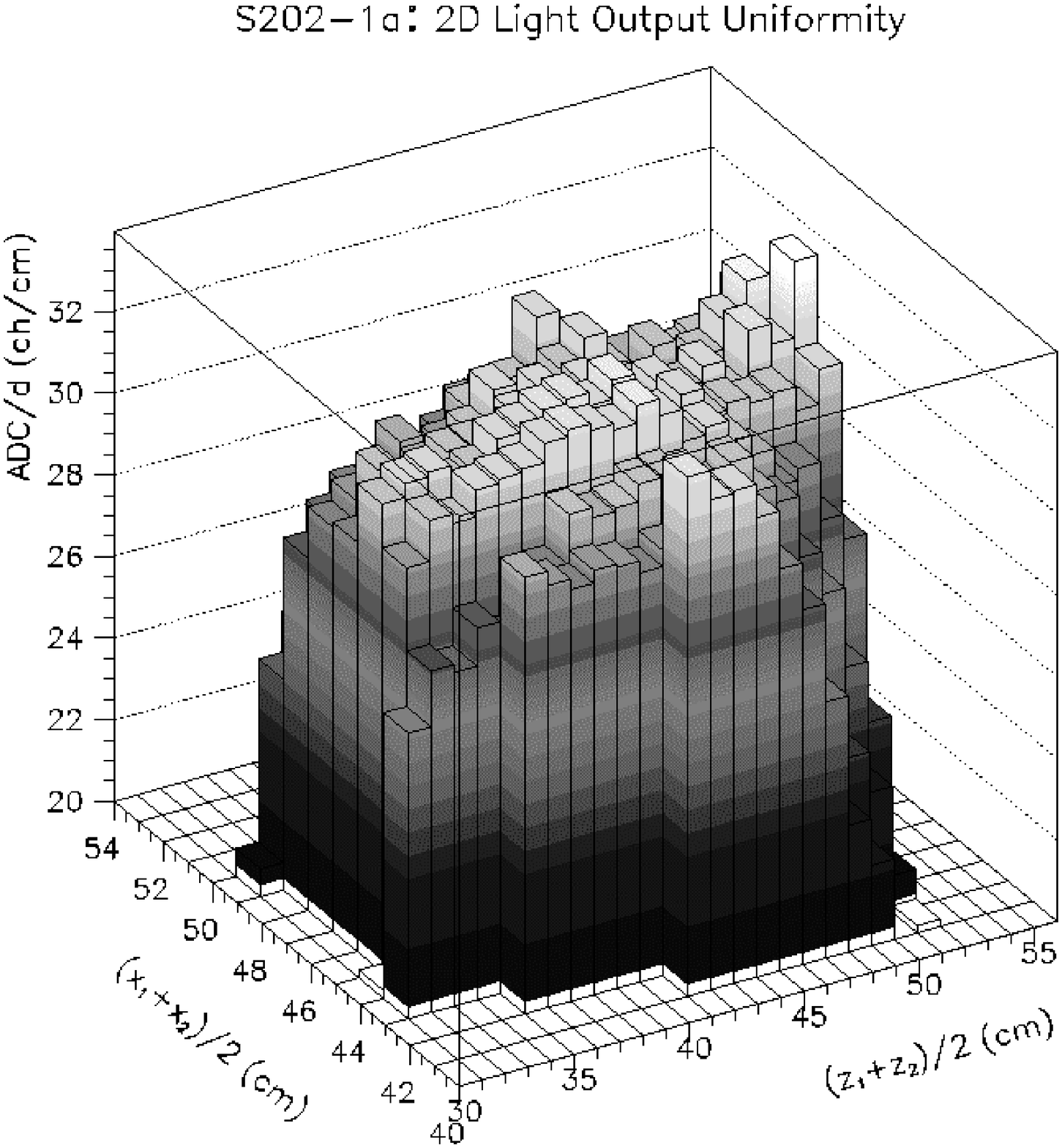,height=11.7cm}}
\vglue 0.25cm
\centerline{FIGURE 10}
\vfill\eject

\centerline{\psfig{figure=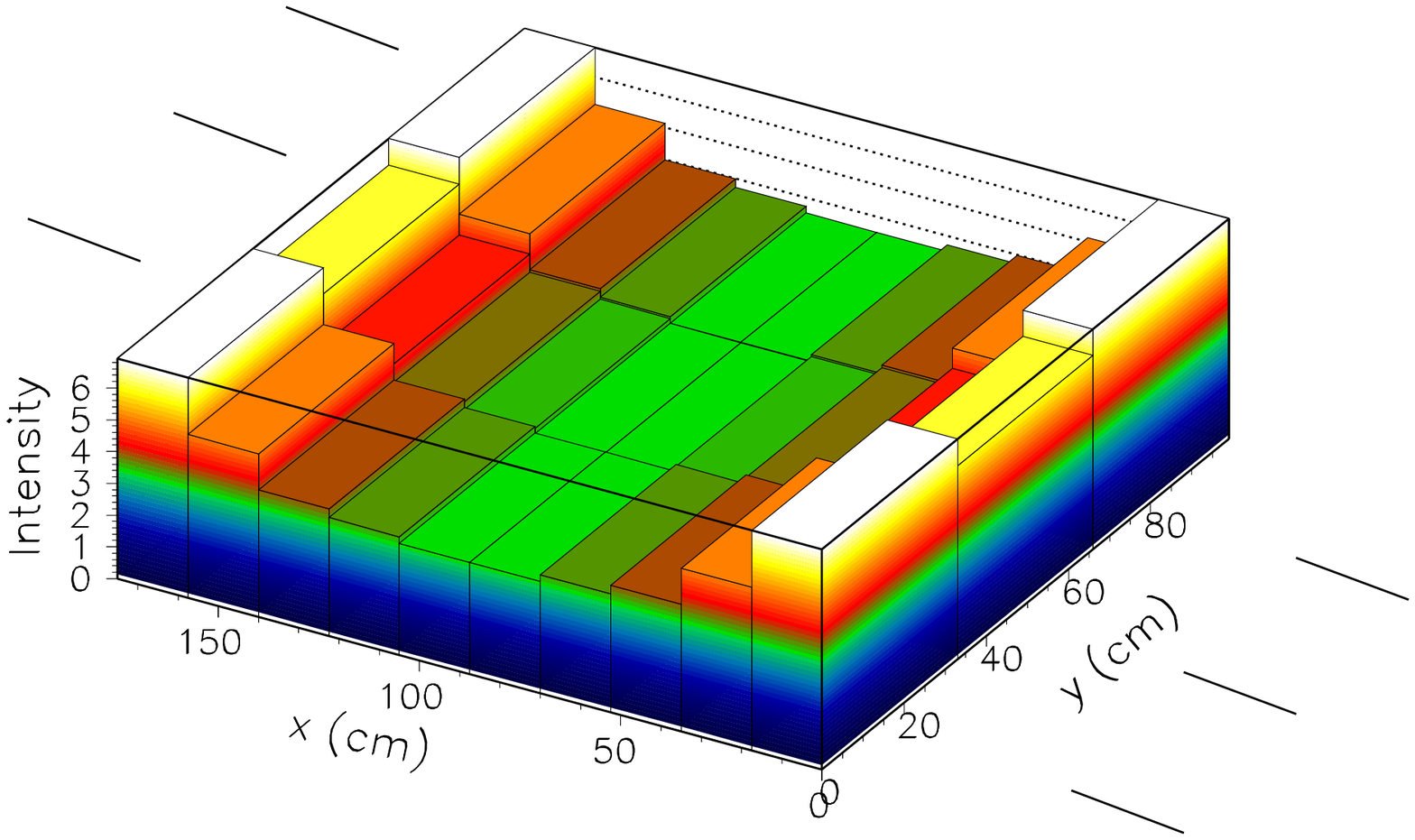,height=18cm}}
\bigskip
\vglue -7.5cm
\centerline{FIGURE 11}
\vfill\eject

\noindent (i)\vglue -1.3cm
\centerline{\psfig{figure=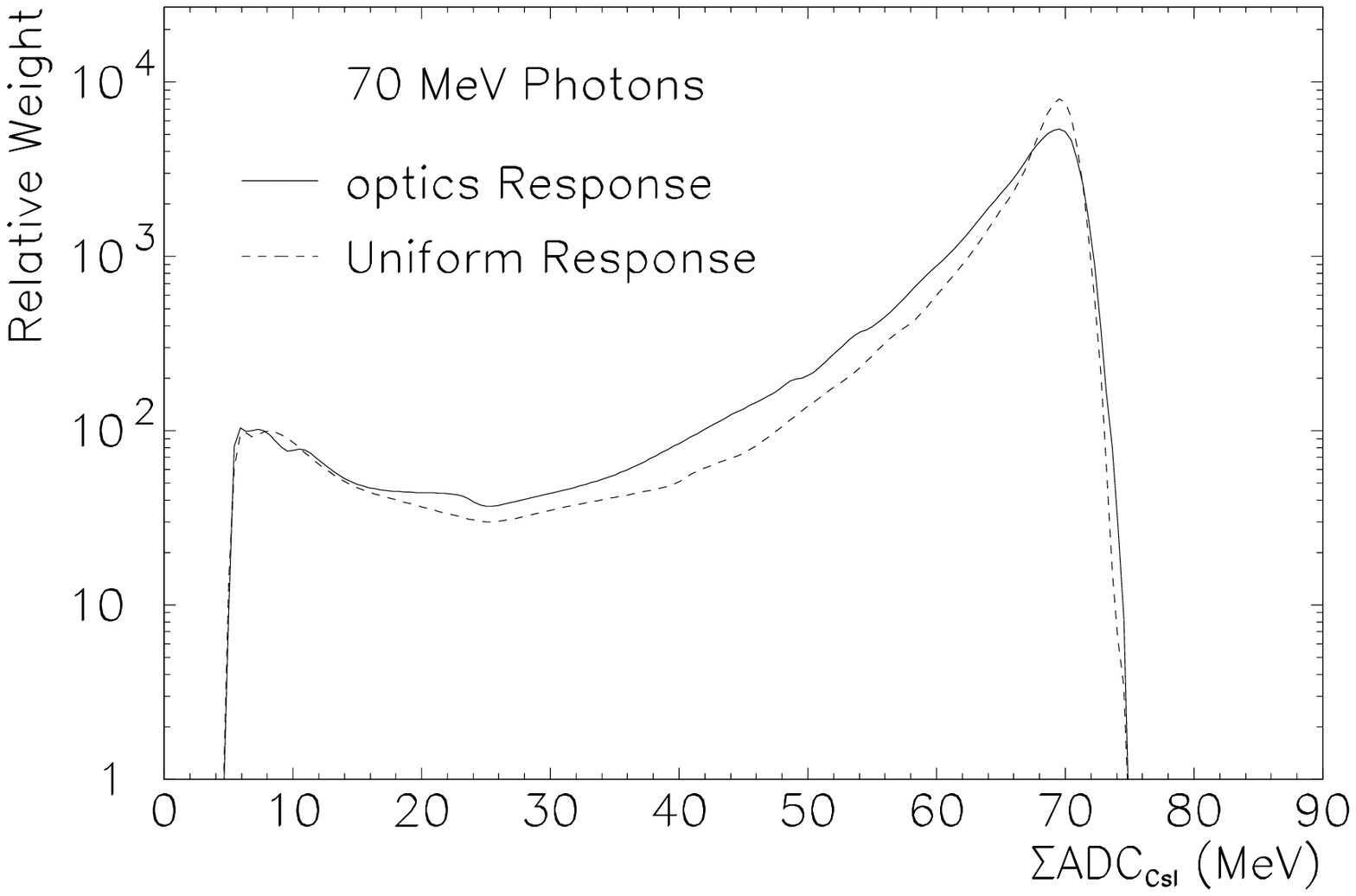,height=18cm}}
\vglue -8.0cm
\noindent (ii)\vglue -0.5cm
\centerline{\psfig{figure=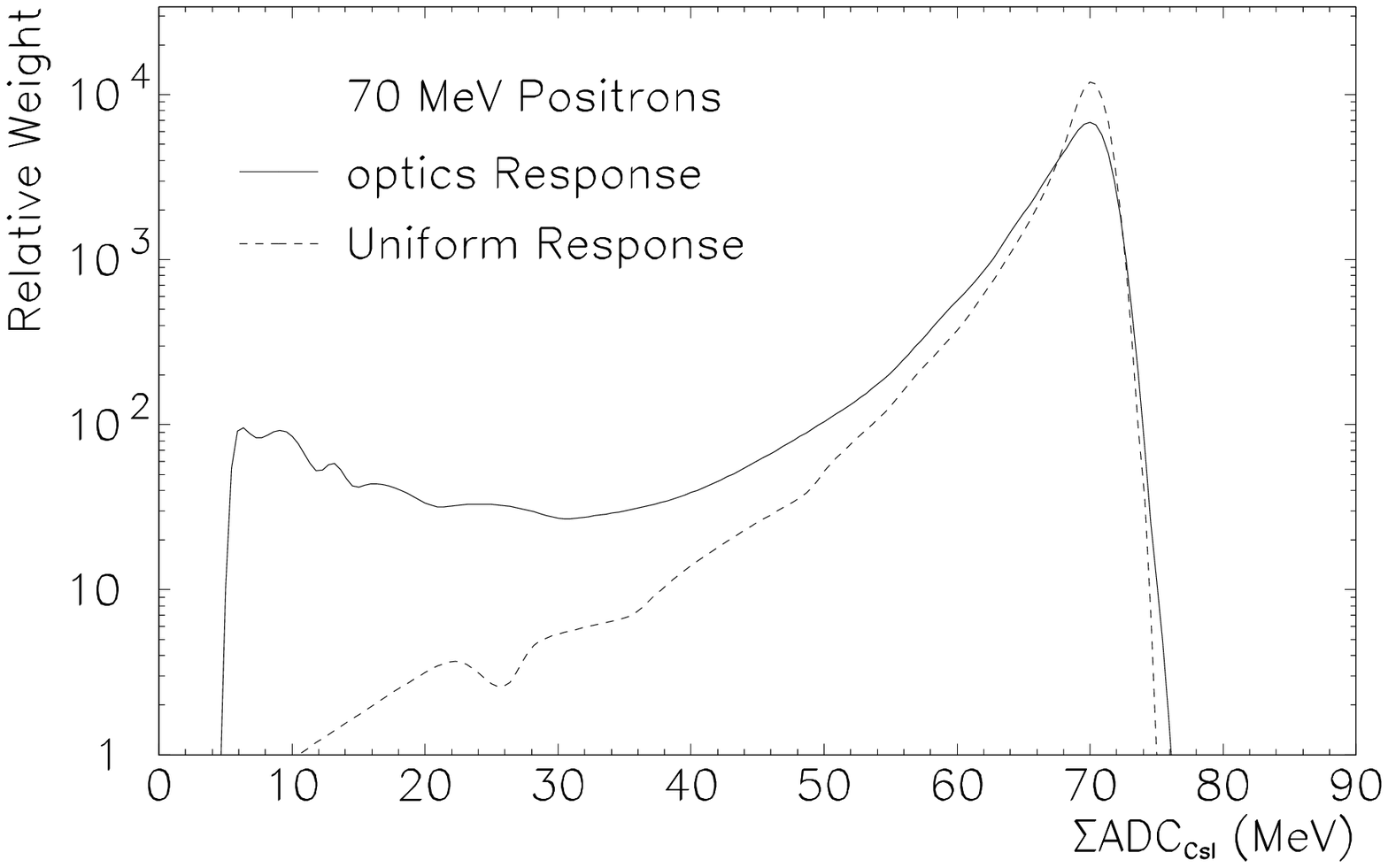,height=18cm}}
\vglue -7.5cm
\centerline{FIGURE 12}
\vfill\eject

}

\end{document}